\documentclass[prl,aps,superscriptaddress,preprint,floatfix]{revtex4-1}

\usepackage{graphicx}
\usepackage{verbatim}
\usepackage{mathrsfs}
\usepackage{amsmath,amsfonts,amssymb}
\usepackage{wrapfig}
\usepackage{bbm}
\usepackage{graphics}

\def\3{2.8in}		
\def\2{2.5in}
\def\4{3.0in}

\newcommand{\sx}{\sigma_1}
\newcommand{\sy}{\sigma_2}

\newcommand{\calc}{{\cal C}}

\newcommand{\calh}{{\cal H}}

\newcommand{\minus}{\text{-}}

\newcommand{\bea}{\begin{eqnarray}}
\newcommand{\enea}{\end{eqnarray}}
\newcommand{\beq}{\begin{equation}}
\newcommand{\eneq}{\end{equation}}

\newcommand{\bpm}{\begin{pmatrix}}
\newcommand{\epm}{\end{pmatrix}}

\newcommand{\bal}{\begin{align}}
\newcommand{\eal}{\end{align}}
\newcommand{\A}{\mathrm{\AA^{\scriptscriptstyle{-1}}}}

\newcommand{\qed}{\nobreak \ifvmode \relax \else
      \ifdim\lastskip<1.5em \hskip-\lastskip
      \hskip1.5em plus0em minus0.5em \fi \nobreak
      \vrule height0.75em width0.5em depth0.25em\fi}

\def \beq {\begin{equation}}
\def \eeq {\end{equation}}
\begin{document}

\title{A new form of (unexpected) Dirac fermions in the \\ strongly-correlated cerium monopnictides}

\author{Nasser~Alidoust$^{\dagger}$} \affiliation {Joseph Henry Laboratory, Department of Physics, Princeton University, Princeton, New Jersey 08544, USA}

\author{A.~Alexandradinata$^{\dagger}$ } \affiliation {Joseph Henry Laboratory, Department of Physics, Princeton University, Princeton, New Jersey 08544, USA} \affiliation {Department of Physics, Yale University, New Haven, Connecticut 06520, USA}

\author{Su-Yang~Xu}\affiliation {Joseph Henry Laboratory, Department of Physics, Princeton University, Princeton, New Jersey 08544, USA}

\author{Ilya~Belopolski}\affiliation {Joseph Henry Laboratory, Department of Physics, Princeton University, Princeton, New Jersey 08544, USA}
\affiliation{Princeton Center for Complex Materials, Princeton Institute for the Science and Technology of Materials, Princeton University, Princeton, New Jersey 08544, USA}

\author{Satya~K.~Kushwaha}\affiliation {Department of Chemistry, Princeton University, Princeton, New Jersey 08544, USA}

\author{Minggang~Zeng}\affiliation {Graphene Research Centre and Department of Physics, National University of Singapore, Singapore 117542}

\author{Madhab~Neupane}\affiliation {Joseph Henry Laboratory, Department of Physics, Princeton University, Princeton, New Jersey 08544, USA}\affiliation{Department of Physics, University of Central Florida, Orlando, FL 32816, USA}

\author{Guang~Bian}\affiliation {Joseph Henry Laboratory, Department of Physics, Princeton University, Princeton, New Jersey 08544, USA}

\author{Chang~Liu}\affiliation {Joseph Henry Laboratory, Department of Physics, Princeton University, Princeton, New Jersey 08544, USA}

\author{Daniel~S.~Sanchez}\affiliation {Joseph Henry Laboratory, Department of Physics, Princeton University, Princeton, New Jersey 08544, USA}

\author{Pavel~P.~Shibayev}\affiliation {Joseph Henry Laboratory, Department of Physics, Princeton University, Princeton, New Jersey 08544, USA}

\author{Hao~Zheng}\affiliation {Joseph Henry Laboratory, Department of Physics, Princeton University, Princeton, New Jersey 08544, USA}

\author{Liang~Fu}\affiliation {Department of Physics, Massachusetts Institute of Technology, Cambridge, Massachusetts 02139, USA}

\author{Arun Bansil}
\affiliation{Department of Physics, Northeastern University, Boston, Massachusetts 02115, USA}

\author{Hsin~Lin}\affiliation {Graphene Research Centre and Department of Physics, National University of Singapore, Singapore 117542}

\author{Robert~J.~Cava}\affiliation {Department of Chemistry, Princeton University, Princeton, New Jersey 08544, USA}

\author{M.~Zahid~Hasan}\affiliation {Joseph Henry Laboratory, Department of Physics, Princeton University, Princeton, New Jersey 08544, USA}
\affiliation{Princeton Center for Complex Materials, Princeton Institute for the Science and Technology of Materials, Princeton University, Princeton, New Jersey 08544, USA}


\begin{abstract}
\textbf{Discovering Dirac fermions with novel properties has become an important front in condensed matter and materials sciences. Here, we report the observation of unusual Dirac fermion states in a strongly-correlated electron setting, which are uniquely distinct from those of graphene and conventional topological insulators. In strongly-correlated cerium monopnictides, we find two sets of highly anisotropic Dirac fermions that interpenetrate each other with negligible hybridization, and show a peculiar four-fold degeneracy where their Dirac nodes overlap. Despite the lack of protection by crystalline or time-reversal symmetries, this four-fold degeneracy is robust across magnetic phase transitions. Comparison of these experimental findings with our theoretical calculations suggests that the observed surface Dirac fermions arise from bulk band inversions at an odd number of high-symmetry points, which is analogous to the band topology which describes a $\mathbb{Z}_{2}$-topological phase. Our findings open up an unprecedented and long-sought-for platform for exploring novel Dirac fermion physics in a strongly-correlated semimetal.}

\end{abstract}
\date{\today}

\maketitle

\bigskip
\noindent\textbf{Introduction}

The search for exotic Dirac fermions with novel functionalities has been a central theme in condensed matter and materials sciences and engineering during the past decade. Following the discovery of graphene \cite{Geim_graphene1} and topological insulators (TIs) \cite{Hasan_TIreview, Qi_TIreview, TI_book_2014, Teo_TI, Hsieh_BiSb, Xia_Bi2Se3, Hsieh_Spin}, many other Dirac systems have followed including topological crystalline insulators (TCIs) \cite{Fu_TCI1, Xu_TCI, Ando_TCI, Story_TCI}, and Dirac \cite{Rappe_DSM, Fang_DSM1, Chen_Na3Bi, Neupane_Cd3As2, Cava_Cd3As2} and Weyl semimetals \cite{Hasan_TaAs1, Dai_TaAs, Hasan_TaAs2, Ding_TaAs}. With the emergence of topological insulators, topologically-protected Dirac surface states in different material systems have improved our fundamental understanding of new quantum phases of matter \cite{Hasan_TIreview, Qi_TIreview}, as well as provided new routes to developing applications in low-power electronics and spintronics devices \cite{Ralph_Torque, Tokura_Hetero, Chen_Tunneling}. Dirac surface states have also been proposed as the missing puzzle for solving some long-standing questions in condensed matter physics, most notably for the Kondo insulators, even though to this date no Dirac fermion band structure has been identified in these materials \cite{Coleman_TKI, Neupane_SmB6, Feng_SmB6, Shi_SmB6}. To date, most of the observed Dirac fermions materialize in weakly-correlated band systems; identification of Dirac fermions in strongly-correlated materials would offer important insight into the intricate physics of these materials. 

Here, we turn our attention to the strongly-correlated cerium monopnictides Ce\textit{X} (\textit{X} = Bi, Sb). We find that Ce$X$ displays some properties that can be associated with a negative-indirect-gap topological insulator. Examining their surface electronic band structure with angle-resolved photoemission spectroscopy (ARPES), we find two highly anisotropic and interpenetrating Dirac fermions with negligibly weak hybridization. Despite the lack of symmetry protection, the observed Dirac fermions manifest a four-fold degeneracy that is robust across magnetic phase transitions -- Ce\textit{X} presents a case study of robustly non-hybridizing Dirac fermions in a topological, strongly-correlated system. Our observations are in sharp contrast with graphene and conventional TIs and TCIs. In those materials, Dirac fermions are separated in momentum space and thus do not hybridize with each other. This does not hold for the two Dirac fermions in Ce\emph{X} as they live at the same location in momentum space. Despite this, they are experimentally as robust as the Dirac cones in conventional TIs and TCIs.

\bigskip
\noindent\textbf{Review of low-carrier, strongly-correlated cerium monopnictides}

In Ce\textit{X} (\textit{X} = Bi, Sb), each Ce atom is trivalent with a singly-occupied 4\textit{f} electron, resulting in a rich variety of Kondo-type behavior \cite{Kasuya1985, Suzuki1995}. This behavior cannot be understood through a single-impurity Kondo model, since there are much fewer carriers than there are magnetic ions; specifically, in CeBi (resp. CeSb), there are 0.021 (resp. 0.029) carriers per Ce ion \cite{Suzuki1993}. These low carrier concentrations also emphasize the role of long-ranged Coulomb interactions between conduction electrons, which essentially stabilize strongly-correlated phases - for these reasons, CeX are classified as low-carrier, strongly-correlated systems \cite{Kasuya1993, Kasuya1994, Yaresko_Correlated}; this class of systems also includes the high-\textit{T}$_\text{c}$ cuprate superconductors \cite{Kasuya1995}.

Ce\emph{X} exhibit a complicated range of magnetically-ordered phases at low temperatures \cite{Yaresko_Correlated}. CeBi transits from a paramagnetic phase to an antiferromagnetic (AFM) type-I phase ($+-$ stacking of ferromagnetic planes) at 25 K, and then changes into another AFM type-IA phase ($++--$ stacking) at 13 K \cite{Kumigashira_CeBi}. Similarly, paramagnetic CeSb transits at 16 K to an antiferroparamagnetic (AFP) phase, which consists of ferromagnetic and paramagnetic Ce(001) layers. Further lowering temperature to 8 K, the AFP phase transforms into a simpler type-IA AFM phase \cite{Takahashi_CeSb}. These low-temperature phases have been interpreted as a magnetic-polaron liquid and lattice, which essentially rely on interactions between 4\textit{f} moments in a Kondo lattice, as well as interactions between conduction electrons \cite{Kasuya1993, Kasuya1994}.

These magnetic transitions are also reflected in their transport properties \cite{Suzuki_Resistivity}. In particular, the logarithmic temperature dependence of the resistivities of these materials exemplifies typical heavy-fermion Kondo systems \cite{Kasuya_Wachter} with a Kondo temperature of about 100 K for CeSb \cite{Kasuya1993}. Other well-documented heavy-fermion behavior includes the correlation-induced enhancement of the electronic effective masses \cite{Suzuki1993}. The large Kerr rotation angles (e.g. 90$^{\circ}$ for cleaved single crystals of CeSb \cite{Wachter_Kerr}) in these materials have also been linked to strong correlation effects \cite{Yaresko_Correlated}. Not all these experimental features are adequately described by density-functional theory,  which motivated several applications of dynamical mean-field theory (DMFT) \cite{Laegsgaard1998, Sakai2005, Sakai2007, Litsarev2012}.

To motivate our work, photoemission and bremsstrahlung isochromat spectroscopies have emerged as sensitive probes of many Kondo-related phenomena \cite{Gunnarsson1983, Gunnarsson_Gschneidner, Lynch_Gschneidner}. Early photoemission studies \cite{Franciosi1981, Gudat_Wachter, Kumigashira_CeBi, Kumigashira_CeP} on Ce\textit{X} (\textit{X} = Bi, Sb) have revealed two peaks in the 4\textit{f} photoemission peaks, at approximately 0.6 eV and 3.0 eV below the Fermi level. These peaks are attributed to hybridization between the induced 4\textit{f} photohole and the conduction 6\textit{p} electrons from the X atoms; this hybridization leads to antibonding and bonding states, which respectively account for the 0.6 and 3.0 eV peaks \cite{Sakai1984, Takeshige1985, Kasuya1985}. These double peaks encode correlation effects through dynamical screening from the intra-atomic \textit{d-f} Coulomb interaction \cite{Takeshige1985, Takeshige1991}. The existing interpretations of the 4\textit{f} double peaks in these materials have thus far relied on the single-impurity, Anderson model \cite{Gunnarsson1983, Bickers1987}, i.e., these works assume that the 4\textit{f} electrons (impurities) on each Ce atom do not mutually interact, and therefore cannot capture momentum-dependent dispersion of the 4\textit{f} emission \cite{Andrews1996, Kumigashira_CeP, Laegsgaard1998} which are especially important in the low-temperature, Kondo-lattice phases \cite{Kasuya1993, Kasuya1994}. This motivates our report of the first momentum-resolved measurement of the 4\textit{f} emission.

\bigskip
\noindent\textbf{Crystal structure}

Ce\textit{X} possesses a rocksalt crystal structure, in which the Ce atoms form a face-centered cubic Bravais lattice, while the pnictogen (\emph{X}) atoms lie on the octahedral voids of this lattice (see Fig. S1 in the Supplementary Information). The bulk Brillouin zone (BZ) is a truncated octahedron with six square faces and eight hexagonal faces; both the bulk BZ and its projection to the (001) surface BZ are shown in Fig. 1\textbf{b}. We have performed X-ray diffraction and core-level measurements on single crystals of CeBi and CeSb, which confirm the high quality and excellent crystallinity of our studied samples (see Figs. S2-S4).

\bigskip
\noindent\textbf{Bulk electronic bandstructure}

While Ce\textit{X} is semimetallic from the perspective of transport, it is instructive to view it as a negative-indirect-gap insulator with the valence band at $\Gamma$ (in the bulk BZ) rising above the Fermi level; this view is supported by our first-principles calculation of the bulk bands as well as our measurements in Fig. 1\textbf{d}. The entire valence band of Ce\textit{X} may then be characterized by topological numbers which distinguish various classes of quantum groundstates \cite{schnyder2008A}; different topological numbers are distinguished by strikingly different surface properties \cite{fidkowski2011}. These numbers can be deduced from our first-principles calculations, which reveal an inverted ordering of the Ce-$d$ and \emph{X}-$p$ orbitals at three symmetry-related $X$ points ($X_1$, $X_2$ and $X_3$  of Fig. 1\textbf{b}). Since non-magnetic Ce\emph{X} is both centrosymmetric and time-reversal symmetric, an odd number of parity inversions implies that the compound is a $\mathbb{Z}_{2}$-topological phase \cite{fu2007a}. The low-energy description of each \textit{X} point is a 3D massive Dirac fermion in the bulk BZ; it is well-known that gapless states localize on the interface between two distinct mass regimes \cite{field4, field5}. 

\bigskip
\noindent\textbf{Surface electronic bandstructure and interpenetrating Dirac cones}

To probe the existence of these gapless surface states, we show in Fig. 1\textbf{a} the ARPES Fermi surface and intensity maps obtained at various binding energies for CeBi. Hole-like pockets are observed at $\bar{\Gamma}$ and we also observe intensities at the $\bar{M}$ point of the surface BZ (these positions are illustrated in the white dashed square of Fig. 1\textbf{a}). A closer analysis of the $\bar{M}$ pockets reveals that their constant energy contours shrink to a point at \textit{E}$_{\text{B}} \simeq 0.25$ eV, and expand at higher binding energies. These pockets are further investigated by a high-resolution zoomed-in Fermi surface in Fig. 1\textbf{c}, which encompasses two $\bar{\Gamma}$ points from the first and second BZs, as well as the two $\bar{M}$ points shared between them. At each equivalent $\bar{M}$ point, the pocket is composed of two interpenetrating ellipses, which center at $\bar{M}$ and extend along the $\bar{\Gamma}-\bar{M}$ direction. As illustrated in Fig. 1\textbf{d}, the ARPES spectra near $\bar{\Gamma}$ and along $\bar{\Gamma}-\bar{M}$ match very well with the bulk bands from our first-principles calculations (overlaid on the ARPES data in the right panel). In the bulk gap at $\bar{M}$, we observe Dirac bands whose absence in the bulk calculation is strongly suggestive of their surface-like character. These bands can be clearly seen in Fig. 1\textbf{d}, and are indeed responsible for forming the elliptical pockets at the Fermi surface. The Fermi surface of CeSb is nearly identical to that of CeBi (see the Fermi surface map in Fig. 2\textbf{a}), and again the bands at $\bar{\Gamma}$ match very well with those predicted from the first-principles calculations (see Figs. S5 and S6).

Let us determine the dispersion of CeSb surface states at $\bar{M}$, by examining two different cuts along the dashed lines shown in Fig. 2\textbf{a}. The evolution of these two Dirac cones in the vicinity of the BZ corner is shown in Fig. 2\textbf{b}. Both cones appear to be gapped away from the $\bar{M}$ point (left panel in Fig. 2\textbf{b}), but merge together at this high-symmetry momentum (right panel in Fig. 2\textbf{b}). Thus, we confirm that the Dirac nodes of both of these cones are positioned right at $\bar{M}$ at the same binding energy of $\simeq 0.40$ eV ($\simeq 0.25$ eV in CeBi). Our high-resolution ARPES $k-E$ cut along $\bar{\Gamma}-\bar{M}$, presented in Fig. 2\textbf{c}, shows these two interpenetrating Dirac cones and their overlapping Dirac nodes at the same momentum location and binding energy. We can also clearly resolve the anisotropy of these Dirac cones in this ARPES spectrum, since one appears as a thin cone inside another elongated cone, which confirms the elliptical shape of these pockets in the Fermi surface maps.

To distinguish between surface and bulk states, we study how the electronic structure  evolves as a function of the incident photon energy ($h\nu$); we expect to probe bulk states of different momentum component $k_z$, while surface states are not expected to evolve. Fig. 3\textbf{a} represents ARPES spectra along the $\bar{\Gamma}-\bar{M}$ direction of the BZ upon varying the incident photon energy. This figure shows that the Dirac cones do not disperse with $h \nu$, thus supporting their surface origin. In contrast, the bands near $\bar{\Gamma}$ and along $\bar{\Gamma}-\bar{M}$ show clear dispersion.

We claim that these interpenetrating Dirac cones originate from bulk band inversions at $X_1$ and $X_2$ (see Fig. 1\textbf{b}). Suppose the inversion at $X_1$ produces one of the elliptical Dirac cones at $\bar{M}$. The rocksalt structure implies that $X_1$ and $X_2$ are related by four-fold rotational symmetry, hence we expect a second elliptical Dirac cone which is rotated from the first by $\pi/2$. Since $X_1$ and $X_2$ project onto two $\bar{M}$ points which are made equivalent by a surface reciprocal vector (see Fig. 1\textbf{b}), these two Dirac cones share a common center. These arguments are consistent with the following effective Hamiltonian for the surface Dirac fermions:

\bal
\calh(k_x,k_y) = \begin{pmatrix} v_1 k_x \sy - v_2 k_y \sx & 0 \\ 0 & v_2 k_x \sy - v_1 k_y \sx \end{pmatrix},
\end{align}
\\
with $\sigma_j$ being Pauli matrices in a pseudospin representation. Our momentum coordinates $(k_x,k_y)$ are chosen relative to the high-symmetry point $\bar{M}$ in the (001) surface BZ, and their directions are parallel to the axes in Fig. 1\textbf{b}. The difference $|v_1|-|v_2|$ is a measure of the anisotropy;  $|v_1|=0.98$ eV$\A$ and $|v_2|=4.48$ eV$\A$ are fitted parameters to the CeSb data, and the resultant dispersion in Fig. 2\textbf{d} has close overlap with the observed cones. In the Supplementary Information, we derive how one flavor of Dirac fermion (upper block of $\calh$) arises from the bulk inversion at $X_1$, while the second flavor originates from $X_2$; this establishes a correspondence between gapless surface states and a topological twist of the bulk wavefunctions.

A peculiarity of the surface Dirac states is that they appear not to hybridize with each other, i.e., we observe no energy gap opening at momentum locations where the cones overlap. Where the nodes of both cones overlap, we then have an intriguing four-fold degeneracy. We note that the symmetry of the ideal (001) surface is that of a square lattice (point group $C_{4v}$ with time-reversal symmetry) \cite{C4v}, and this symmetry group does not protect four-fold degeneracies \cite{kosterbook}. While symmetry-allowed hybridizations can in principle remove this degeneracy, we find that our minimal model ($\calh$) of two unhybridized Dirac fermions works remarkably well in reproducing our measurements. Our assumption of $C_{4v}$ symmetry is supported by the symmetrical shape of the observed Dirac cones, e.g., each cone reflects into itself, and a $\pi/2$ rotation relates one cone to the other. We remark that this symmetry is preserved under surface relaxation and rumpling, and future investigations of rumpling effects might shed some light on this four-fold degeneracy.

Despite not being protected by symmetry, this degeneracy is surprisingly robust in a few different ways. (i) Even though CeBi and CeSb have different material parameters, they both share this `accidental' degeneracy. (ii) This degeneracy persists even in the magnetic phases, where time-reversal symmetry is spontaneously broken. In the temperature-dependent measurements of CeSb (Fig. 3\textbf{b}), we find that the Dirac cones remain intact across the paramagnetic-AFP-AFM transitions. The corresponding data for CeBi can be found in Fig. S7, where again the Dirac surface states remain unchanged as the high-temperature paramagnetic phase transits to the two low-temperature AFM phases. These measurements clearly indicate the robustness of these Dirac surface states across the various magnetic phase transitions in CeBi and CeSb.

\bigskip
\noindent\textbf{4\textit{f} final-state emission}

For CeSb, Fig.\ 4\textbf{a} and \textbf{b} show respectively the momentum-resolved photoemission at the 4\emph{f}-resonant \cite{Franciosi1981,Allen,Gudat_Wachter} photon energy of 122 eV and the off-resonant 128 eV; their difference isolates emission from the 4\emph{f} bands, as shown in Fig.\ 4\textbf{c}; the momentum-integrated difference emission in Fig.\ 4\textbf{d} agrees well with earlier photoemission studies \cite{Gudat_Wachter}.  Here, the 4\textit{f} flat bands are related to final-state emission \cite{Lynch_Gschneidner}, while the interpenetrating Dirac cones are attributed to single-particle, initial states; their simultaneous observation marks the first identification of topological surface states in a low-carrier, strongly-correlated system.

It has been suggested for CeP that the momentum-dependence of the 4\textit{f} dispersion arises from \textit{p-f} mixing around $\Gamma$ and intra-atomic \textit{d-f} mixing around $X$ \cite{Kumigashira_CeP}. For CeBi and CeSb, this momentum-dependent mixing indirectly follows from the topological band inversion, which changes the bulk orbital character from Bi-6\textit{p} (at $\Gamma$) to Ce-5\textit{d} (at $X$). A rigorous calculation to support this hypothesis is still lacking, and future work will shed light on whether the other members of the cerium monopnictides (CeP, CeN and CeAs) also display topological surface states.

\bigskip
\noindent\textbf{Topological characterization in the nonmagnetic phase}

Our surface analysis has thus far focused on two interpenetrating Dirac cones. Given that our first-principles calculations predict non-magnetic Ce\textit{X} to be a $\mathbb{Z}_{2}$-topological phase, we might expect to see an odd number of surface Dirac cones \cite{Hasan_TIreview}. Indeed, given bulk inversions at $X_1$ and $X_2$, the symmetries of the rocksalt structure dictate that a similar bulk inversion occurs at $X_3$. This last inversion naively leads to a third surface Dirac cone at $\bar{\Gamma}$ with approximately the same energy as the other two cones. However, in this energy range we instead observe a bulk continuum of bands, as supported by our first-principles calculation (see Fig. 1\textbf{d}). The hybridization between this third cone and the bulk continuum may be strong enough to delocalize the cone, and explains its lack of experimental signatures. In contrast, the double cones at $\bar{M}$ are energetically separated from any bulk state. We therefore have a negative-indirect-gap TI with effectively an even number of surface Dirac fermions.

While our discussion of non-magnetic Ce\emph{X} so far has centered on topological properties protected by time-reversal symmetry,
a complete characterization must also account for its crystalline symmetries \cite{Teo_TI, Fu_TCI1}. Indeed, the surface Dirac cones of non-magnetic Ce\emph{X} lie over a plane (indicated by the purple plane in the bulk BZ of Fig. 1\textbf{b}) which is invariant under the  reflection: $y \rightarrow -y$; in short, we call this a mirror plane. Bloch states on this mirror plane may be distinguished by whether they are odd or even under this reflection. In Ce\emph{X}, the even and odd subspaces both exhibit a quantum anomalous Hall effect but with opposite chiralities; this is characterized by an integer invariant ($\calc_+$) called the mirror Chern number \cite{Teo_TI}. As we show in the Supplementary Information, two distinct phases may arise from bulk inversions at the $X$ points: if the parameters $v_1$ and $v_2$ in our effective Hamiltonian have the same (resp. opposite) sign, then $\calc_+=+1$ (resp. $-3$). While the absolute values of $v_1$ and $v_2$ may be determined from the measured energy dispersions, their relative sign does not affect the energies, but is instead encoded in the spin texture of the surface states.

Ce\emph{X} may be instructively compared with the rocksalt family of SnTe insulators \cite{Fu_TCI1}, which have a mirror Chern number of $-2$ but are trivial under the classification by time-reversal symmetry. For SnTe, band inversions at two inequivalent $L$ points project to the same $\bar{X}$ point in the (001) BZ. A field-theoretic study in Ref. \onlinecite{field5} also predicts two Dirac cones at $\bar{X}$; in comparison, our two Dirac cones lie at $\bar{M}$. For SnTe, it was argued that additional lattice effects lead to hybridization of the Dirac cones, as has been observed experimentally \cite{Xu_TCI}; these lattice effects correspond to large-momentum scattering between the two flavors of Dirac fermions. In this perspective, Ce\emph{X} presents a counter-example where lattice effects are seemingly irrelevant and a field-theoretic description is sufficient.

\bigskip
\noindent\textbf{Summary and outlook}

We find an unprecedented type of Dirac fermions in the cerium monopnictides, which is uniquely distinct from those of graphene and previously discovered topological insulators, as we schematically illustrate in Fig. 5\textbf{b}-\textbf{d}. For each monopnictide, our measurements clearly resolve  two anisotropic Dirac fermions at the corner of the surface Brillouin zone. These Dirac fermions appear not to hybridize with each other, and exhibit an intriguing four-fold degeneracy where the nodes of both cones overlap; this degeneracy is not protected by the symmetries of the material. Given a parity inversion of the bulk bands at each $X$ point, we have analytically derived an effective Hamiltonian of these surface Dirac cones which overlaps closely with the ARPES data. Our first-principles slab calculation also reproduces the observed surface states upon fine-tuning the surface potential, as shown in Fig. 5\textbf{a} (resp. Fig. S9) for the case of CeSb (resp. CeBi). While this fine-tuning was necessary to produce the four-fold degeneracy in our first-principles calculation, experimentally the Dirac cones remain robustly degenerate across various magnetic phase transitions and despite our attempts at surface modification with potassium deposition. This `accidental' degeneracy persists for both CeSb and CeBi, which have different material parameters -- one begins to wonder if there is an underlying explanation behind this `accident'.

The observed Dirac fermions are consistent with Ce\emph{X} being a negative-indirect-gap topological insulator in the time-reversal-symmetric classification. We further predict a third surface Dirac cone centered at $\bar{\Gamma}$ which is masked by bulk bands; in future work one can envision applying pressure or tuning the alloy composition to unmask this third cone. In addition, spin measurements of the surface states would conclusively determine the topology of Ce\emph{X} under crystalline symmetries. 

Finally, our newly-discovered surface Dirac fermions motivate a re-interpretation of previous, extensive studies\cite{Lynch_Gschneidner,Kasuya1985,Kasuya1993,Yaresko_Correlated} of cerium monopnictides and similar correlated materials to account for the role of these Dirac fermions. The robust non-hybridizing nature of our surface Dirac fermions in a strongly-correlated material system opens a new research frontier in condensed matter and materials sciences and engineering.

\bigskip
\noindent\textbf{Methods}

\textbf{Sample growth and electronic structure measurements.} The high-quality single crystals of CeBi and CeSb, with \textit{Fm-3m} structure, were grown respectively from Bi- and Sb-self fluxes. ARPES measurements were performed with incident photon energies of 30 - 100 eV at beamlines 4.0.3 and 10.0.1 of the Advanced Light Source (ALS) at the Lawrence Berkeley National Laboratory (LBNL), and with incident photon energies of 8 - 30 eV at beamline 5-4 of the Stanford Synchrotron Radiation Lightsource (SSRL) at the SLAC National Accelerator Laboratory. Samples were cleaved \textit{in situ} at 10 K in chamber pressure better than $5\times10^{-11}$ torr at both the SSRL and the ALS, resulting in shiny surfaces. Energy resolution was better than 15 meV and momentum resolution was better than 1\% of the surface BZ.

\textbf{First-principles calculation methods.} Our first-principles calculations are performed in the Vienna \textit{ab initio} simulation package (VASP) using the generalized gradient approximation (GGA) and the projector augmented wave (PAW) method \cite{Hafner_DFT, Ernzerhof_DFT, Blochl_DFT}. The geometry optimization of bulk CeBi and CeSb crystals is performed with force convergence criteria at 0.01 eV\AA$^{-1}$. A Monkhorst-Pack k-mesh ($12 \times 12 \times 12$) is used to sample the bulk Brillouin zone \cite{Pack_DFT}. A slab model with CeBi (CeSb) thickness around 6 nm and vacuum thickness larger than 15 \AA \space is adopted to simulate CeBi (CeSb) (001) thin films.  Trivalent Ce potential with \textit{f}-electrons treating as core electrons are adopted. Spin-orbit coupling (SOC) effect is included as a second variational step using eigenfunctions from scalar relativistic calculation \cite{Bechstedt_DFT}.

\bigskip
\noindent\textbf{Acknowledgments}

The work at Princeton and Princeton-led synchrotron-based ARPES measurements are supported by U.S. Department of Energy grant no. DE-FG-02-05ER46200 and U.S. National Science Foundation grant no. NSF-DMR-1006492. Crystal growth was supported by the Army Research Office Multidisciplinary University Research Initiative on topological insulators, grant no. W911NF-12-1-0461. A.A. was supported by NSF CAREER DMR-095242, ONR - N00014\text{-}11\text{-}1-0635, MURI\text{-}130 \text{-}6082, NSF-MRSEC DMR-0819860, Packard Foundation, Keck grant, DARPA under SPAWAR Grant no. N66001\text{-}11\text{-}1-4110 and by the Yale Prize Fellowship. H.L. acknowledges the Singapore National Research Foundation (NRF) for support under NRF award no. NRF-NRFF2013-03. We gratefully acknowledge Jonathan D. Denlinger, Sung-Kwan Mo, and Makoto Hashimoto for technical beamline assistance at the beamlines 4.0.3 and 10.0.1 of the ALS at LBNL, and at beamline 5-4 of the SSRL at the SLAC. S.K.K. acknowledges Jason W. Krizan for discussions about the crystal growth. We also thank B. Andrei Bernevig, Timothy Hsieh, Chen Fang, Xi Dai, and Leonid Glazman for discussions.

\bigskip
\noindent\textbf{Author contributions}

N.A. and M.Z.H conceived and designed the experiments. N.A. performed the experiments with assistance from S.-Y.X., I.B., M.N., G.B., C.L., D.S.S., P.P.S., and H.Z.; A.A. performed theoretical model calculations and related analysis. S.K.K. and R.J.C. prepared and provided samples and performed sample characterization. M.Z., A.B., and H.L. performed first-principles band structure calculations. L.F. suggested the theory of materials class. N.A., A.A., L.F., and M.Z.H. performed data analysis, figure planning, and draft preparation. M.Z.H. was responsible for the overall direction, planning, and integration among different research units.

$^{\dagger}$ These authors contributed equally to this work.

\begin{figure*}
\centering
\includegraphics[width=16.5cm]{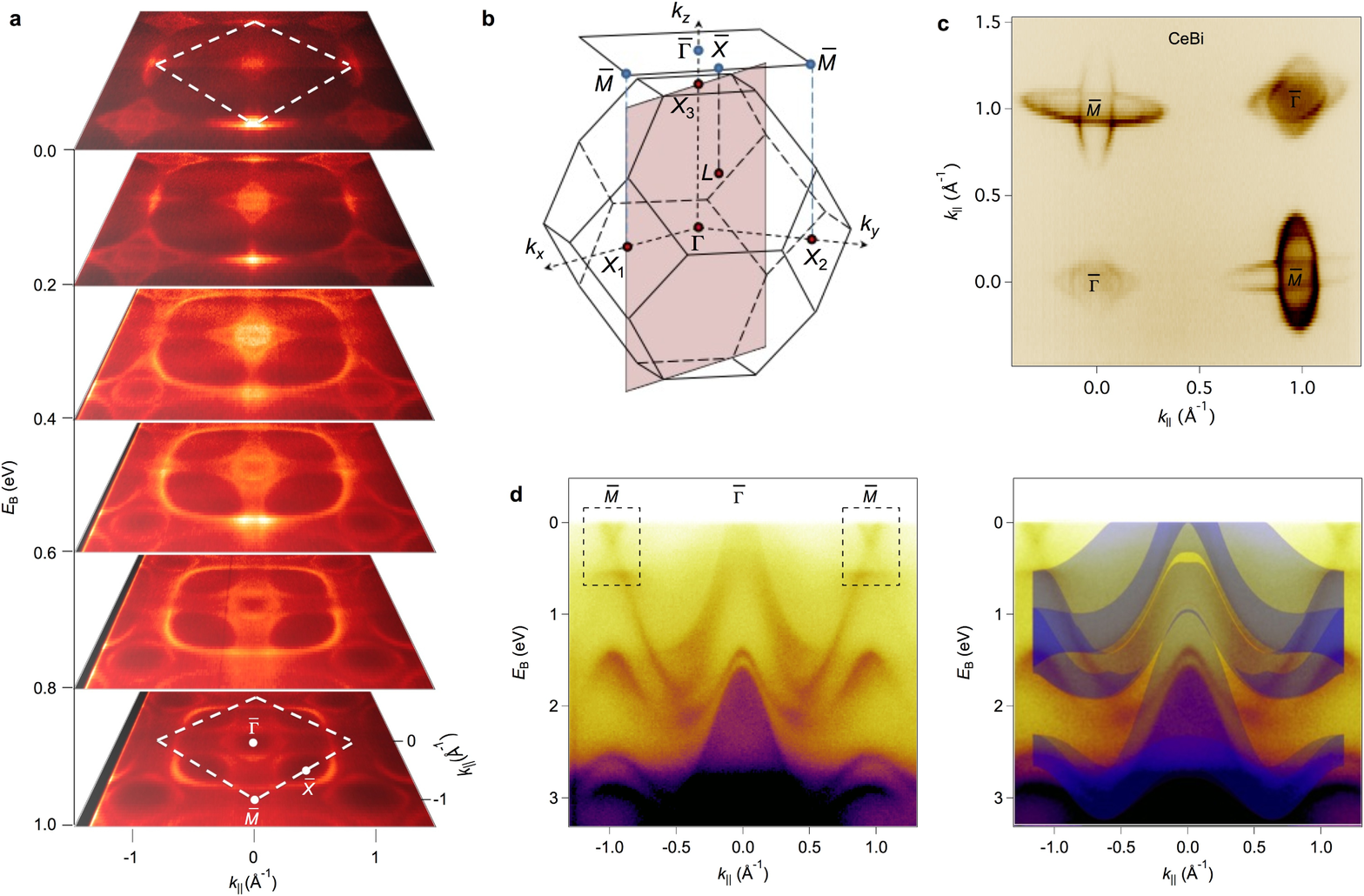}
\caption{\textbf{Observation of Dirac bands in Cerium monopnictides.} \textbf{a}, ARPES Fermi surface map and constant binding energy contours on the (001) cleaving plane of the band structure of CeBi at various energies. \textbf{b}, Brillouin zone (BZ) of cerium monopnictides CeBi and CeSb, and its projection to the (001) surface. The high-symmetry momenta are labeled. The purple plane indicates the mirror plane at $k_y = 0$ in the 3D BZ. \textbf{c}, A zoomed-in Fermi surface map showing the existence of electronic states around the $\bar{\Gamma}$ and the $\bar{M}$ points of the surface BZ. \textbf{d}, ARPES spectra along the high-symmetry direction of $\bar{M}-\bar{\Gamma}-\bar{M}$, and its comparison with the first-principles bulk band structure calculations (overlaid on the ARPES data in the right panel). The observed band structure along this direction matches very well with the calculated bulk bands in the vicinity of $\bar{\Gamma}$ and between $\bar{\Gamma}$ and $\bar{M}$; we thus interpret the Fermi surface in \textbf{c} as arising from hole-like bulk bands at $\bar{\Gamma}$, in addition to Dirac cones at $\bar{M}$ which are not found in the bulk calculation; this strongly suggests the surface-like character of the cones.}
\end{figure*}

\begin{figure*}
\centering
\includegraphics[width=15.5cm]{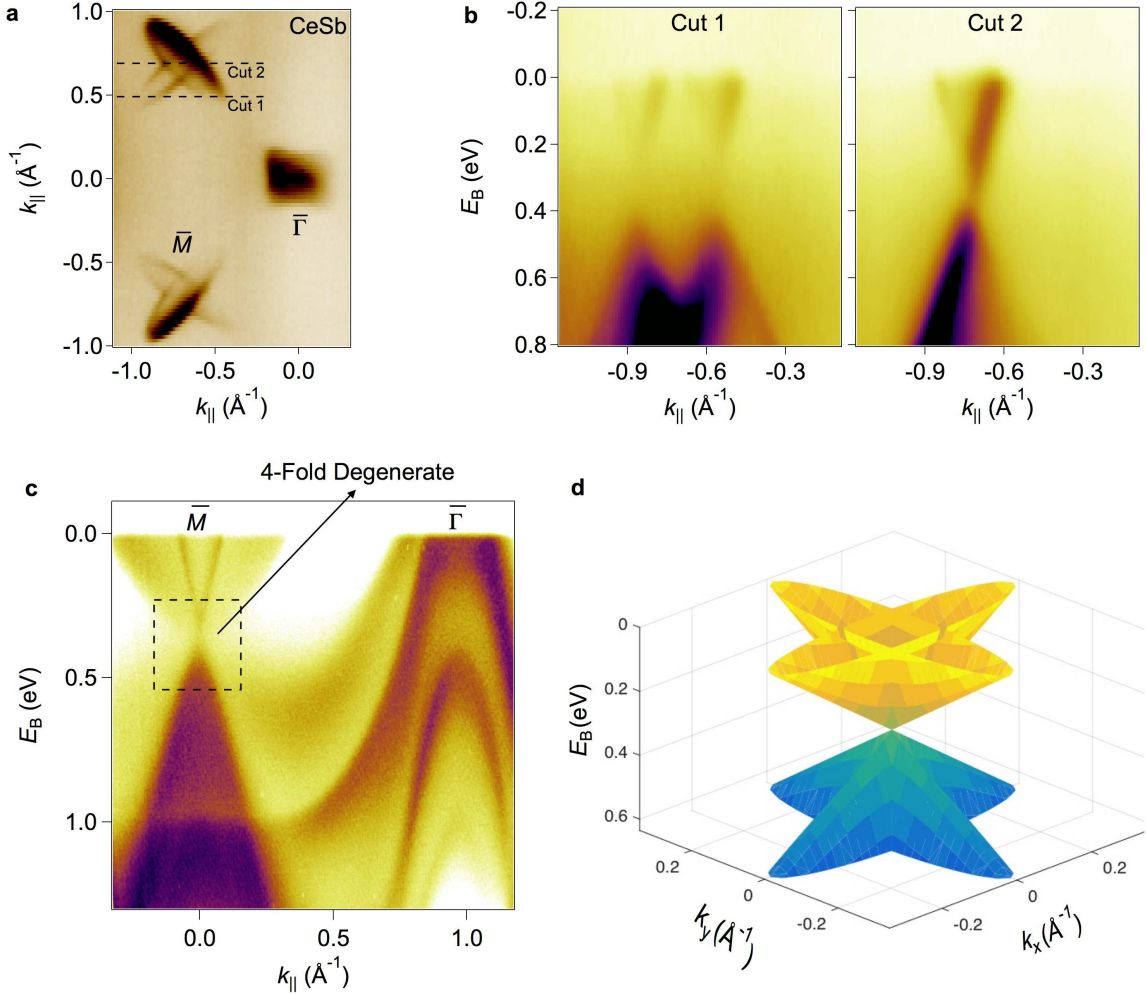}
\caption{\textbf{Anisotropic, interpenetrating, and non-hybridizing Dirac cones.} \textbf{a}, CeSb's Fermi surface is composed of hole-like bands at $\bar{\Gamma}$ and Dirac bands at $\bar{M}$, just as with CeBi. \textbf{b}, $k-E$ cuts taken around $\bar{M}$, along the directions indicated by dashed lines in \textbf{a}. These cuts clearly show a four-fold degeneracy at $\bar{M}$. \textbf{c}, High-resolution ARPES spectra of CeSb along $\bar{\Gamma}-\bar{M}$, which show two anisotropic, interpenetrating Dirac bands near $\bar{M}$ and a four-fold degenerate Dirac point. \textbf{d}, The dispersion of an effective Hamiltonian fitted to the ARPES data, showing close overlap with the experimentally observed Dirac cones.}
\end{figure*}

\begin{figure*}
\centering
\includegraphics[width=16.5cm]{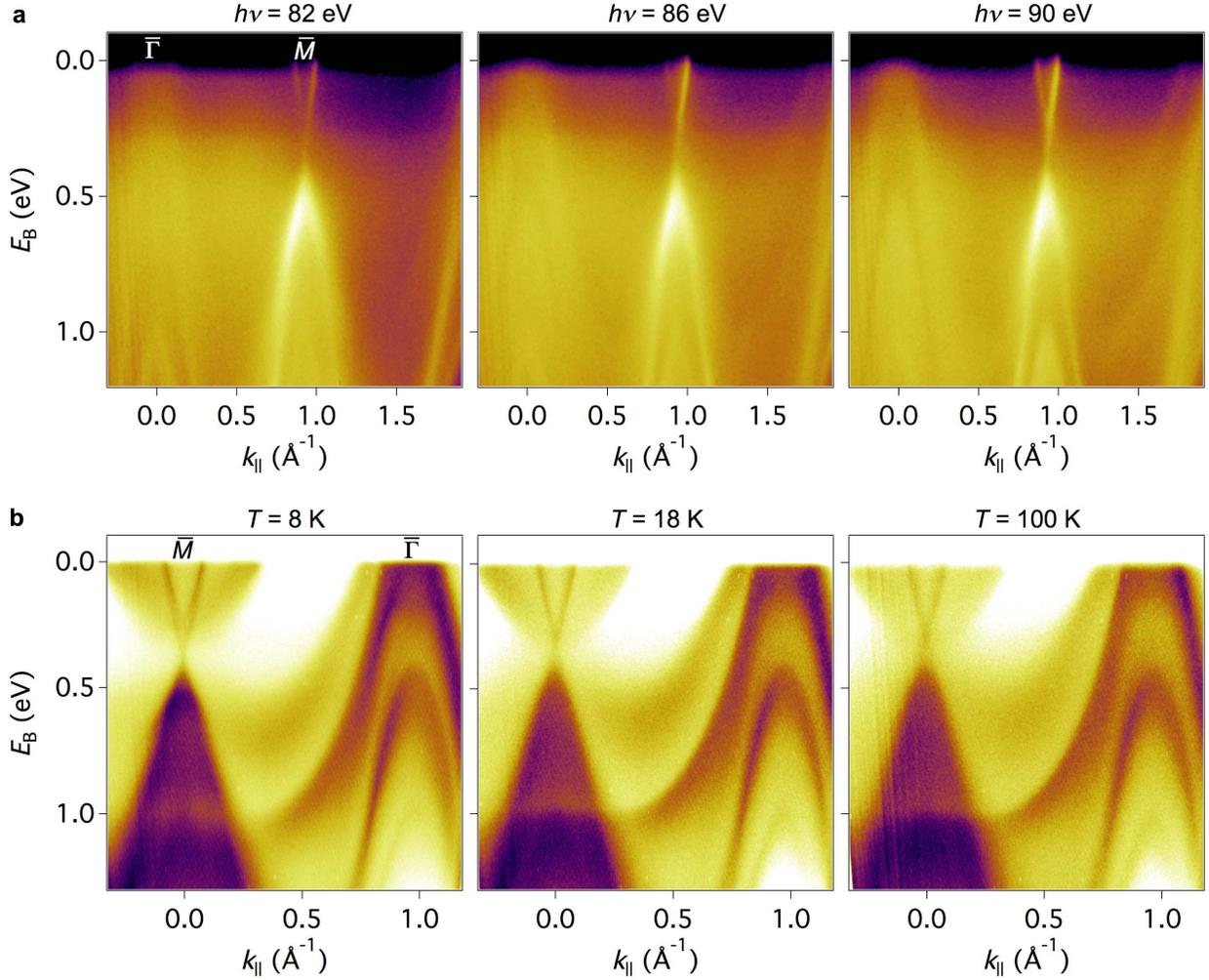}
\caption{\textbf{Surface origin and magnetic insensitivity of the Dirac states.} \textbf{a}, ARPES spectra taken at successive incident photon energies $h\nu$ along the $\bar{\Gamma}-\bar{M}$ direction, to probe different $k_{\text{z}}$. We observe that the Dirac bands at $\bar{M}$ appear to be independent of incident photon energy and thus possess no dispersion along $k_{z}$, revealing that they are indeed surface states. On the other hand, the bands around $\bar{\Gamma}$ and higher biding energies around $\bar{M}$ clearly disperse upon varying $h\nu$, a signature of their bulk origin. \textbf{b}, Temperature-dependent measurements of the CeSb surface states. The Dirac cones are observed to be robust in all three magnetic phases (including the low-temperature antiferromagnetic and antiferroparamagnetic phases and one high-temperature paramagnetic phase).}
\end{figure*}

\begin{figure*}
\centering
\includegraphics[width=15cm]{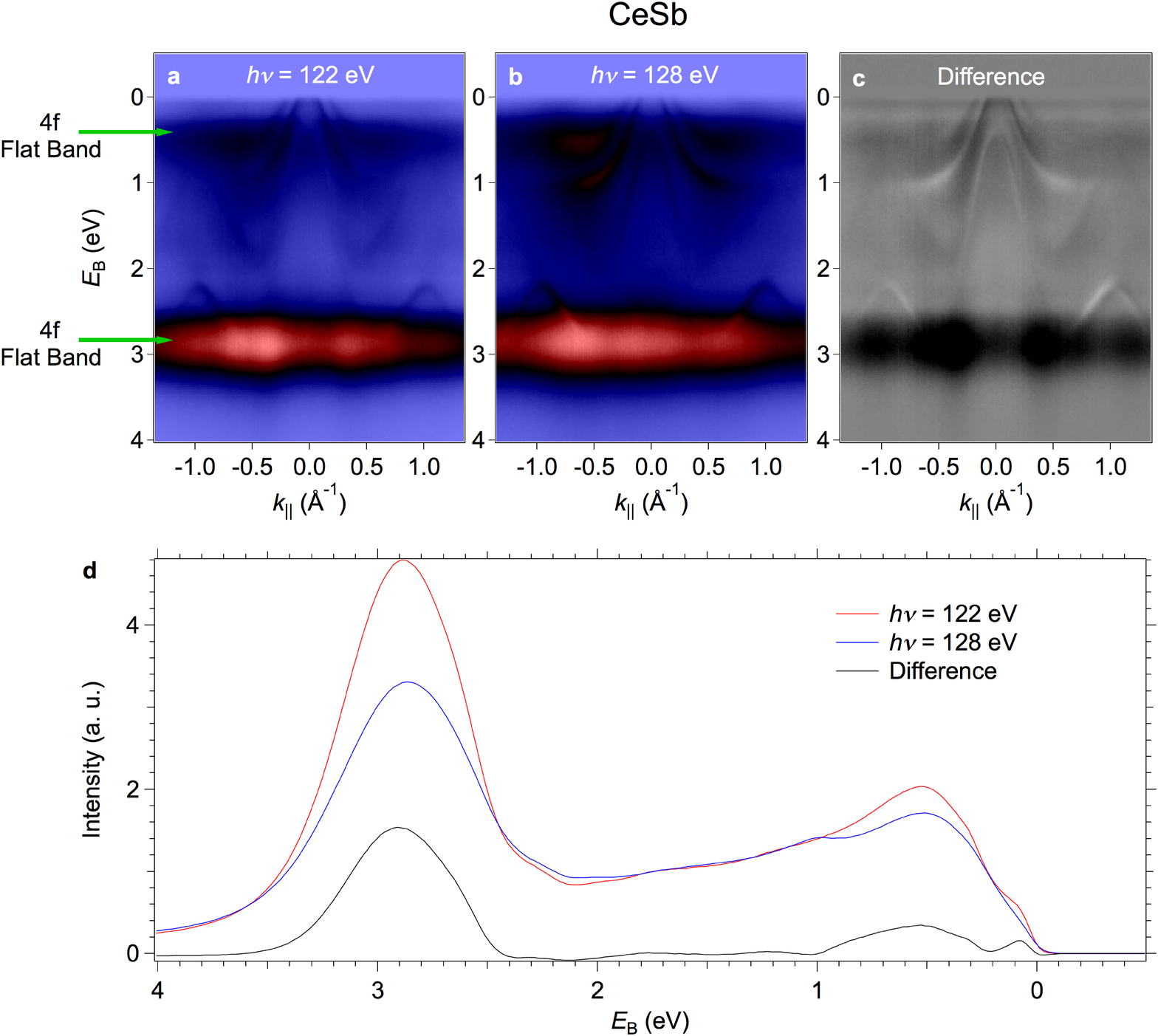}
\caption{\textbf{4\textit{f} flat bands in CeSb.} \textbf{a}, ARPES spectra of CeSb around the $\bar{\Gamma}$ point of the Brillouin zone at photon energy $h\nu$ = 122 eV, showing the existence of a 4\textit{f} flat band close to the Fermi level ($\sim$ 0.5 eV) and another 4\textit{f} flat band at deeper binding energies ($\sim$ 2.9 eV). \textbf{b}, same as \textbf{a} obtained at $h\nu$ = 128 eV. \textbf{c}, The difference of the two spectra in \textbf{a} and \textbf{b}. \textbf{d}, Momentum-integrated energy distribution curves (EDCs) at the photon energies of 122 eV and 128 eV and their difference.}
\end{figure*}

\begin{figure*}
\centering
\includegraphics[width=16.5cm]{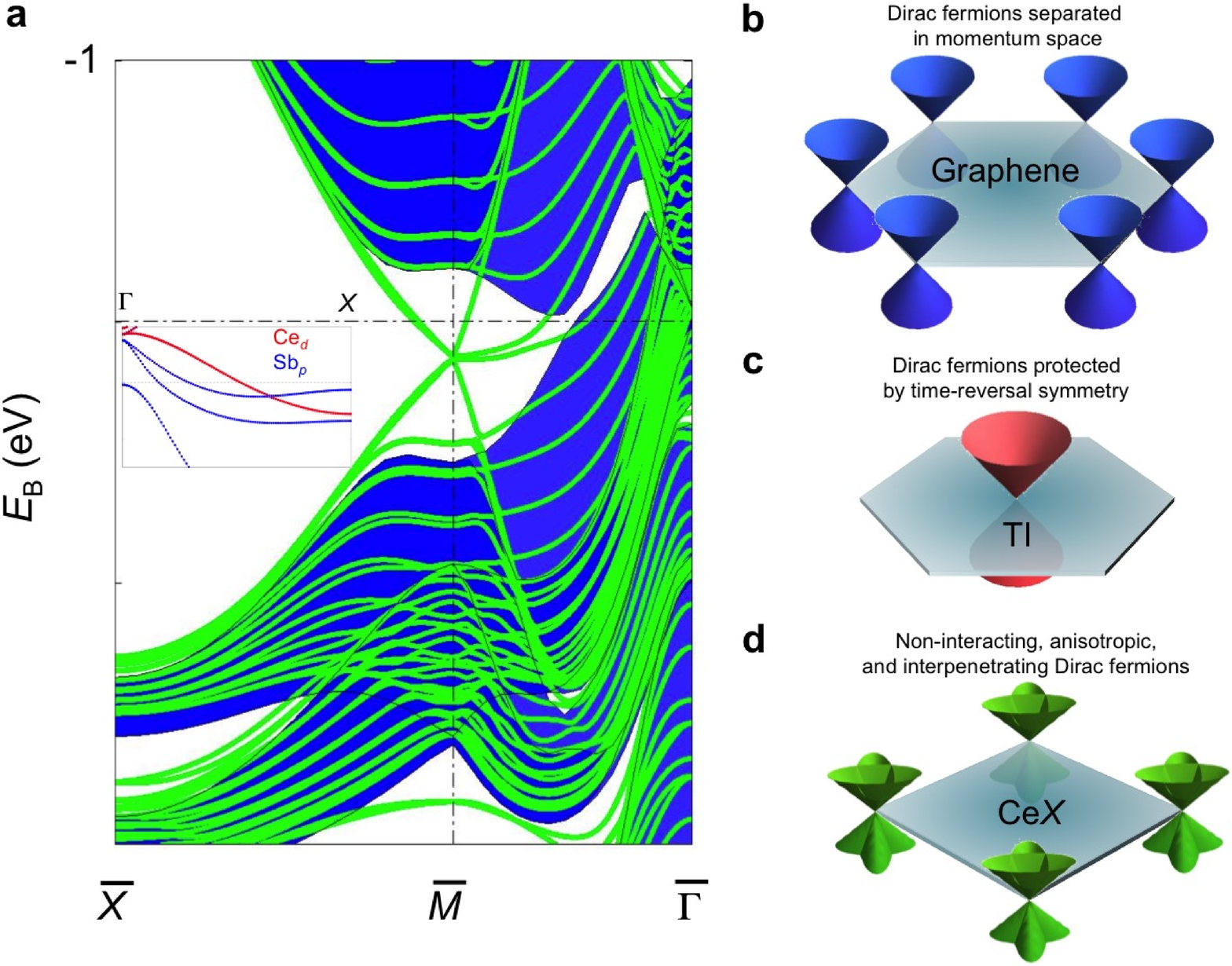}
\caption{\textbf{Dirac surface states beyond graphene and conventional topological insulators.} \textbf{a}, Green lines: first-principles calculation of a CeSb slab; the surface potential on this slab has been fine-tuned to achieve the four-fold degeneracy that is observed experimentally. Blue shaded areas correspond to bulk bands (from a first-principles bulk calculation) projected onto the surface BZ. The inset shows the inversion of Ce-$d$ and Sb-$p$ bands at the $X$ point of the bulk BZ. \textbf{b-d}, The Dirac cones of cerium monopnictides differ from those of graphene and bismuth-based topological insulators in that: (i) the Dirac bands interpenetrate with negligibly weak hybridization and exhibit four-fold degeneracy, which is not protected by symmetry, (ii) there are an even number of Dirac cones, despite our identification of Ce\textit{X} as a $\mathbb{Z}_{2}$-topological phase.}
\end{figure*}

\end{document}


\noindent\textbf{Organization of Supplemental Material}

\noi{\ref{app:cryst}}  Crystal structure, XRD measurements and core levels.

\noi{\ref{app:band}}  Bulk bandstructure calculations and parity analysis.

\noi{\ref{app:temp}}  Dirac cones and $4f$ flat bands in CeBi.

\noi{\ref{app:slab}} Slab bandstructure calculations.

\noi{\ref{app:eff}} Effective Hamiltonian of the surface Dirac fermions.

\noi{\ref{app:poss}} Possible topologies under crystalline symmetries.

\section{C\lowercase{rystal structure and} XRD \lowercase{measurements of cerium monopnictides}}\la{app:cryst}

\begin{figure}[H]
\centering
\includegraphics[width=8cm]{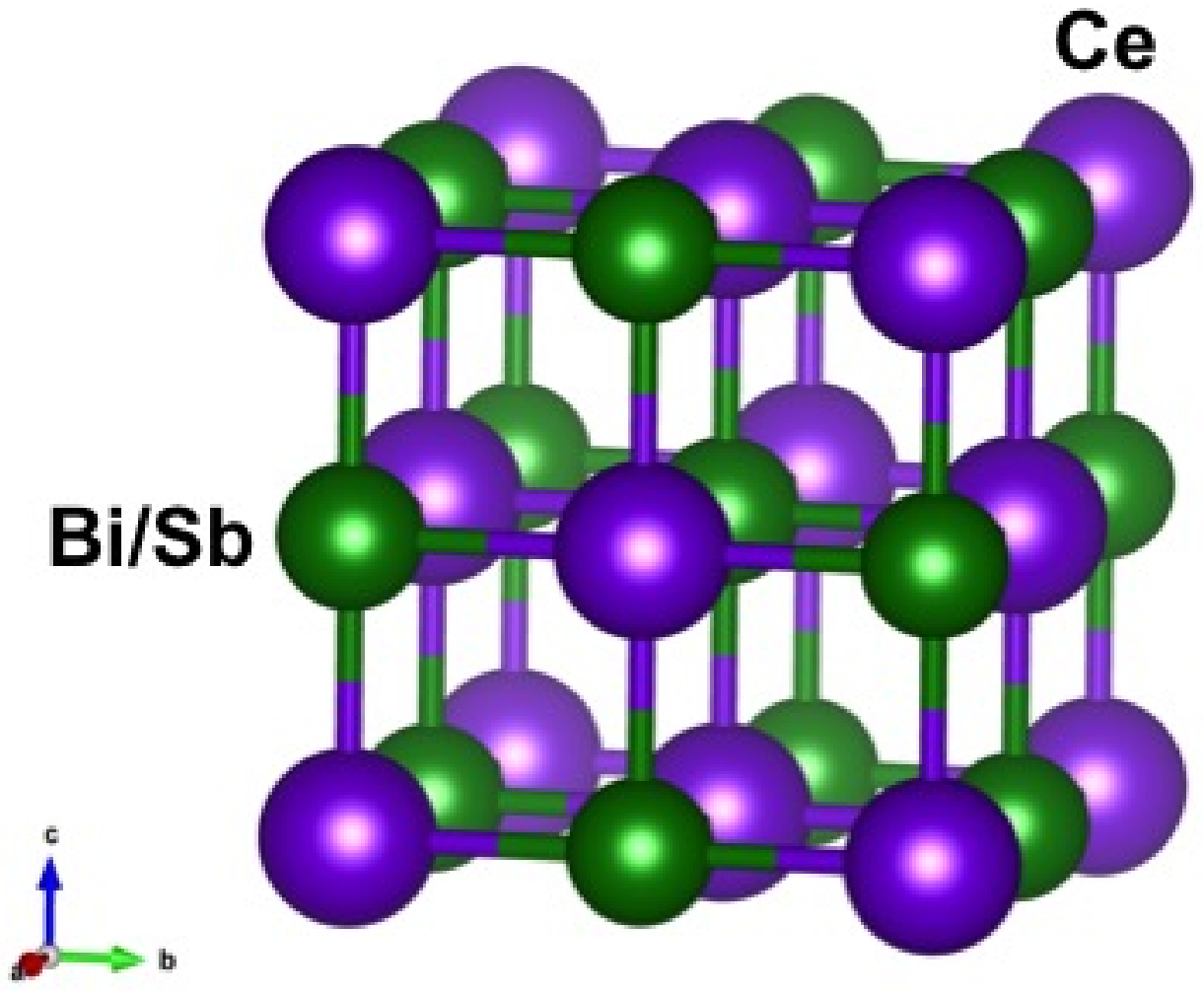}
\caption{\textbf{Crystal structure of Cerium monopnictides.} Ce\textit{X} (\textit{X} = Bi, Sb) possess a rock-salt crystal structure, in which the Ce atoms form a face-centered cubic (fcc) Bravais lattice, while the pnictogen atoms lie on the octahedral voids of this lattice}
\end{figure}

\begin{figure}[H]
\centering
\includegraphics[width=10cm]{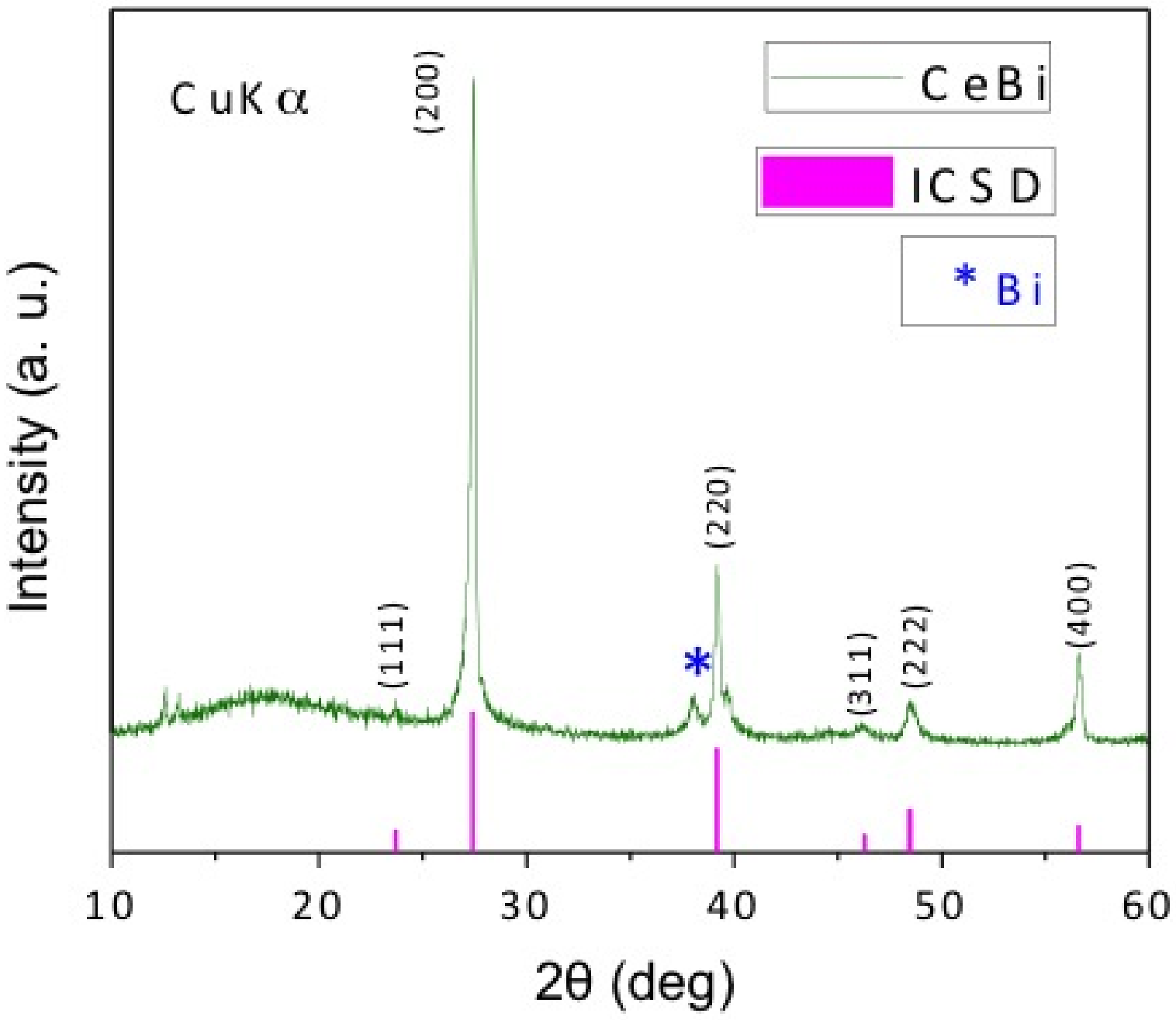}
\caption{\textbf{XRD measurements of CeBi.} X-ray diffraction (XRD) measurements of the CeBi crystals used in our experiments. This data matches well with the reported lattice parameters of CeBi $a = b = c = 6.5$ $\text{\AA}$ from the Inorganic Crystal Structure Database (ICSD).}
\end{figure}

\begin{figure}[H]
\centering
\includegraphics[width=10cm]{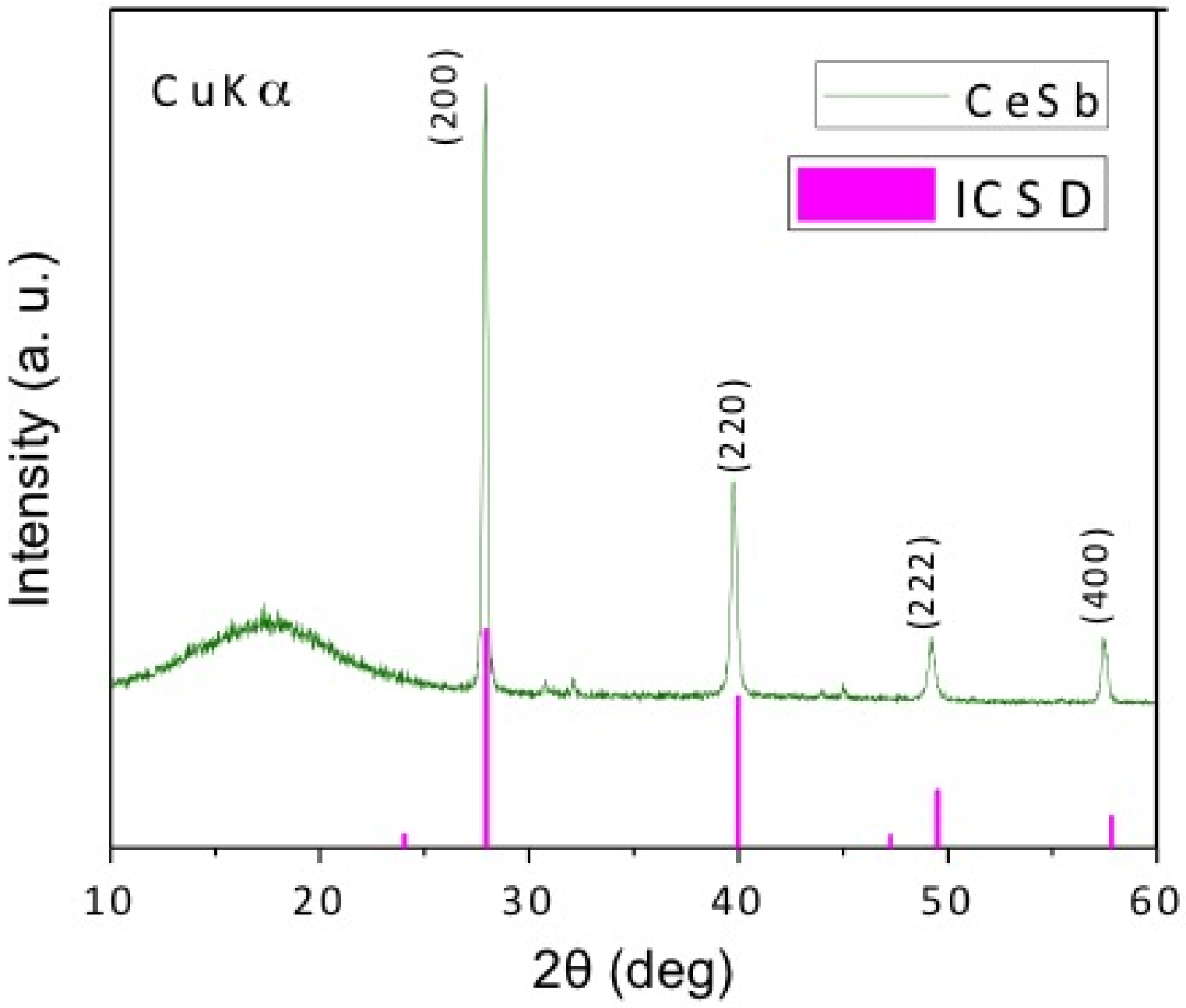}
\caption{\textbf{XRD measurements of CeSb.} X-ray diffraction (XRD) measurements of the CeSb crystals used in our experiments. This data matches well with the reported lattice parameters of CeSb $a = b = c = 6.42$  $\text{\AA}$ from the Inorganic Crystal Structure Database (ICSD).}
\end{figure}

\begin{figure}[H]
\centering
\includegraphics[width=12cm]{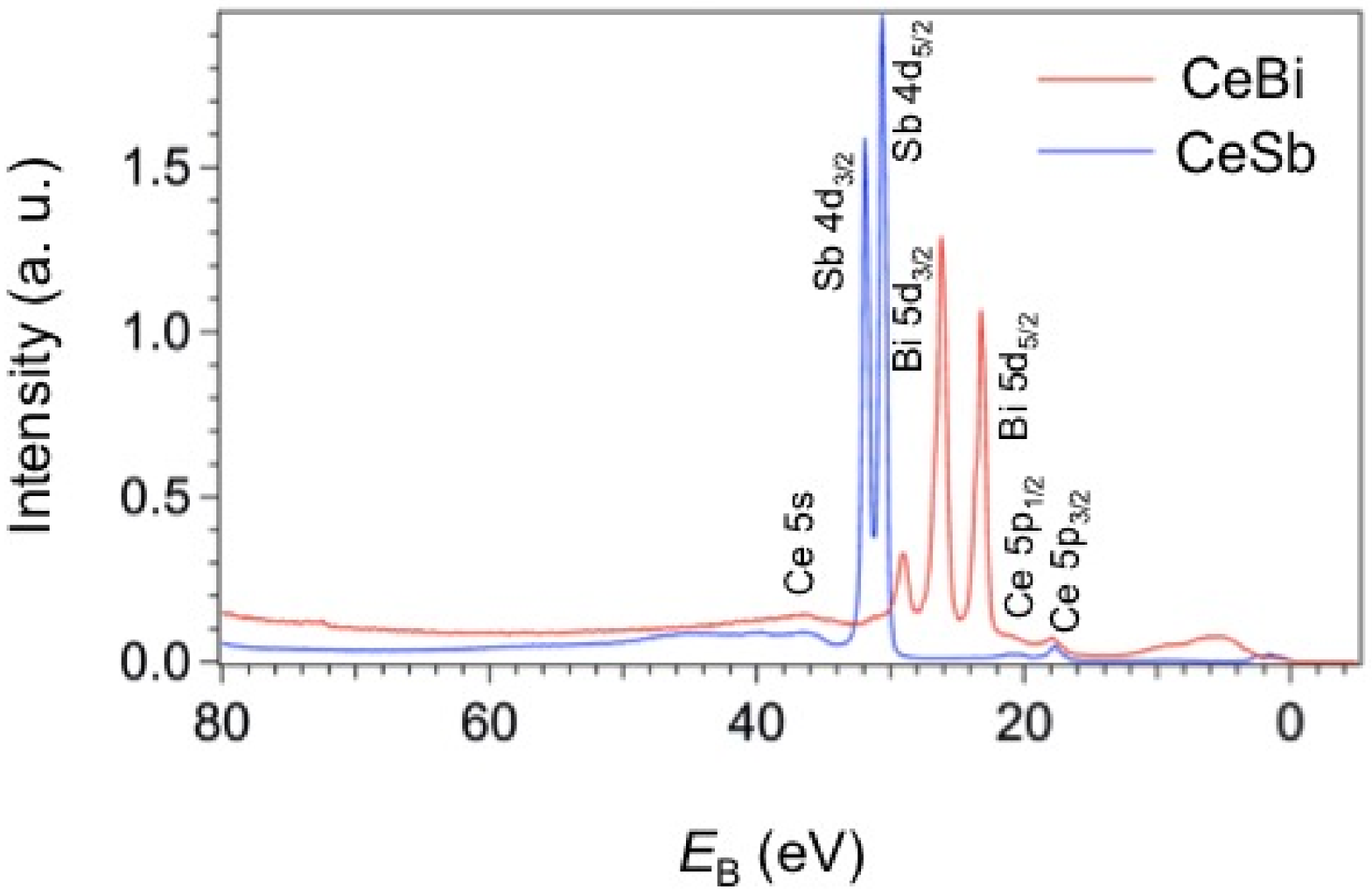}
\caption{\textbf{Core-levels of CeBi and CeSb.} Core-level spectra obtained from the CeBi and CeSb crystals used in our studies. Cerium 5s and 5p, bismuth 5d, and antimony 4d core levels are clearly resolved in our measurements.}
\end{figure}

\section{B\lowercase{ulk bandstructure calculations and parity analysis}}\la{app:band}

\begin{figure}[H]
\centering
\includegraphics[width=12cm]{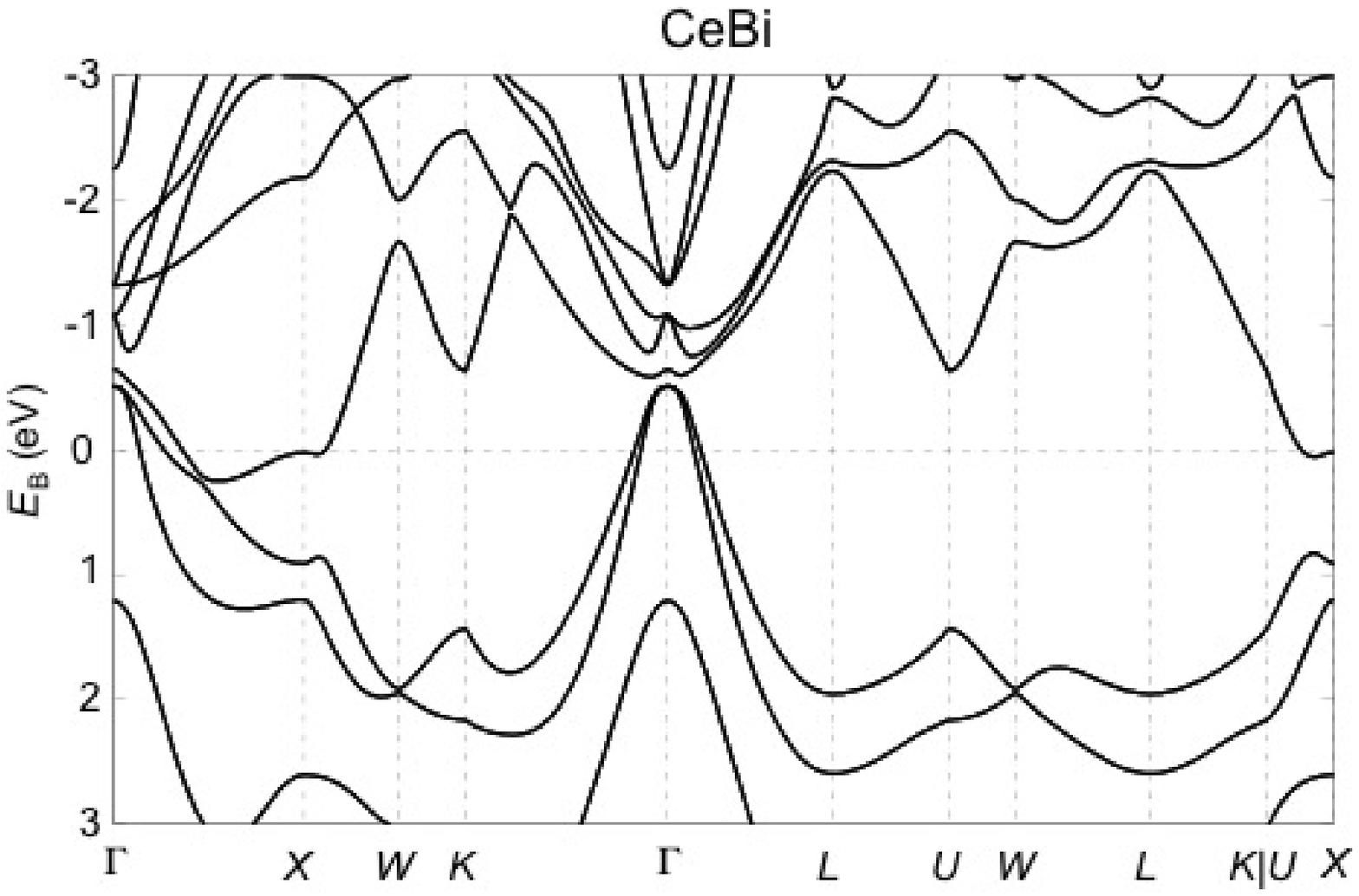}
\caption{\textbf{Bulk bandstructure of CeBi.} First-principles band structure calculations of CeBi. The horizontal dashed line corresponds to 
the Fermi level.}
\end{figure}

\begin{figure}[H]
\centering
\includegraphics[width=12cm]{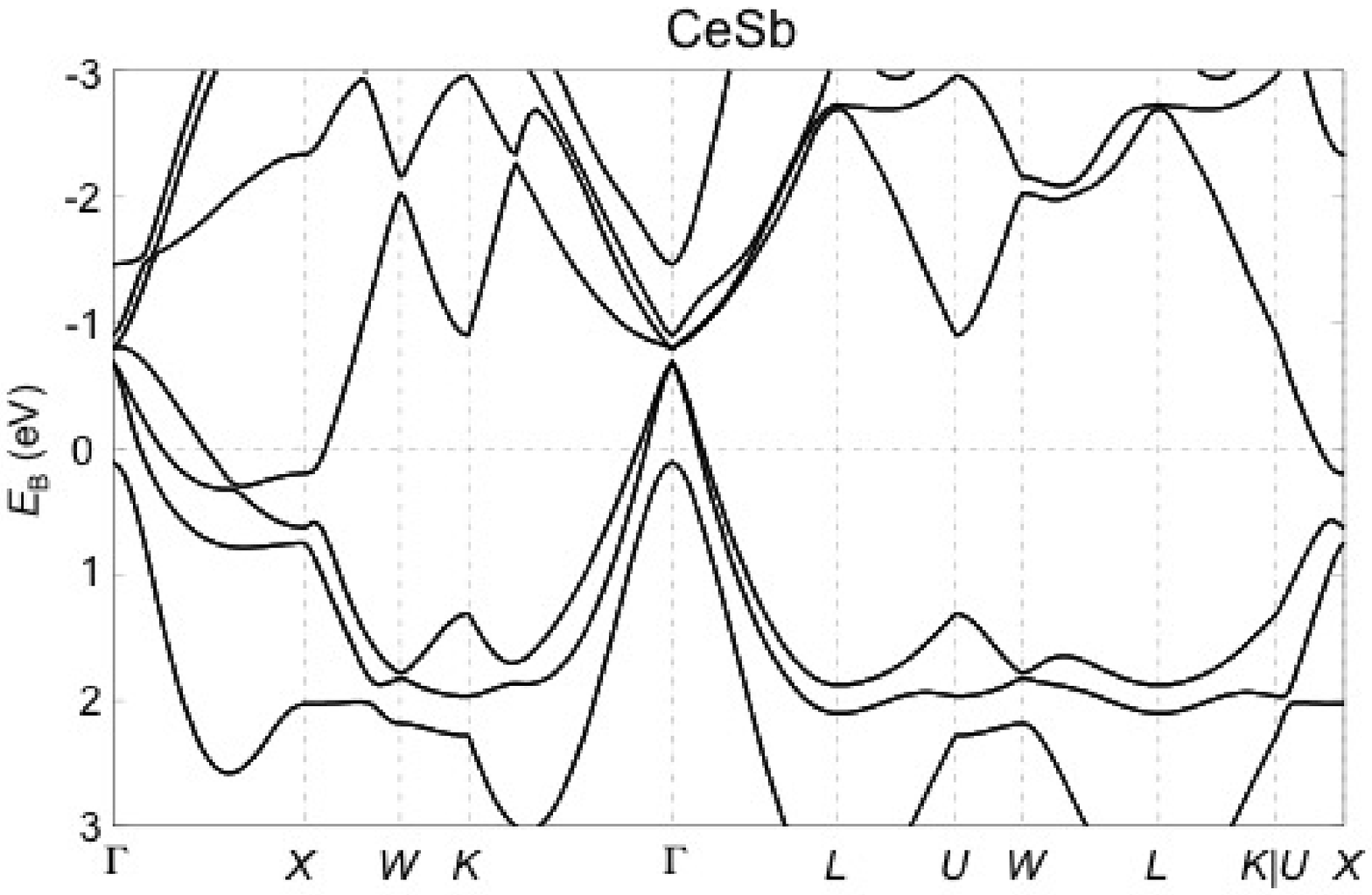}
\caption{\textbf{Bulk bandstructure of CeSb.} First-principles band structure calculations of CeSb. The horizontal dashed line corresponds to the Fermi level.}
\end{figure}

\begin{table}[H]
\centering
\begin{tabular}{|c|c|c|} \hline
TRIM points & Parity of occupied bands & Parity of unoccupied band  \\ \hline
  $1\Gamma$ & $---$ & $-$ \\
  $3X$ & $--+$ & $-$ \\
  $4L$ & $---$ & $+$ \\
  \hline
\end{tabular}
\caption{The parity eigenvalues of the three highest occupied bands and  the first lowest conduction band of CeBi at eight TRIM points. The product of the occupied parity eigenvalues is $-1$, indicating the nontrivial Z2 topology of CeBi.}
\end{table}

\section{D\lowercase{irac cones and $4f$ flat bands in} C\lowercase{e}B\lowercase{i}}\la{app:temp}

\begin{figure}[H]
\centering
\includegraphics[width=14cm]{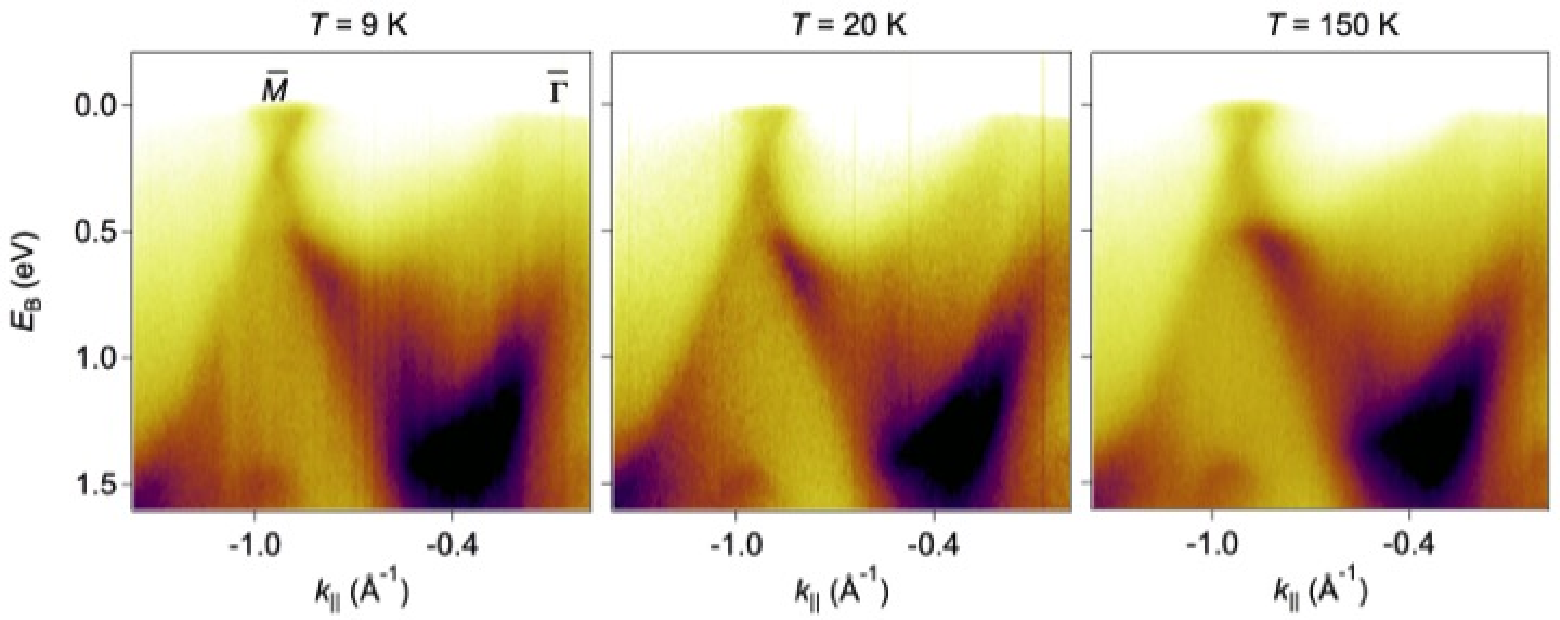}
\caption{\textbf{Temperature dependence of the Dirac cones in CeBi.} The Dirac cones are shown to be robust in both low-temperature (type-IA and type-I) AFM phases and also in the high-temperature paramagnetic phase.}
\end{figure}

\begin{figure}[H]
\centering
\includegraphics[width=15cm]{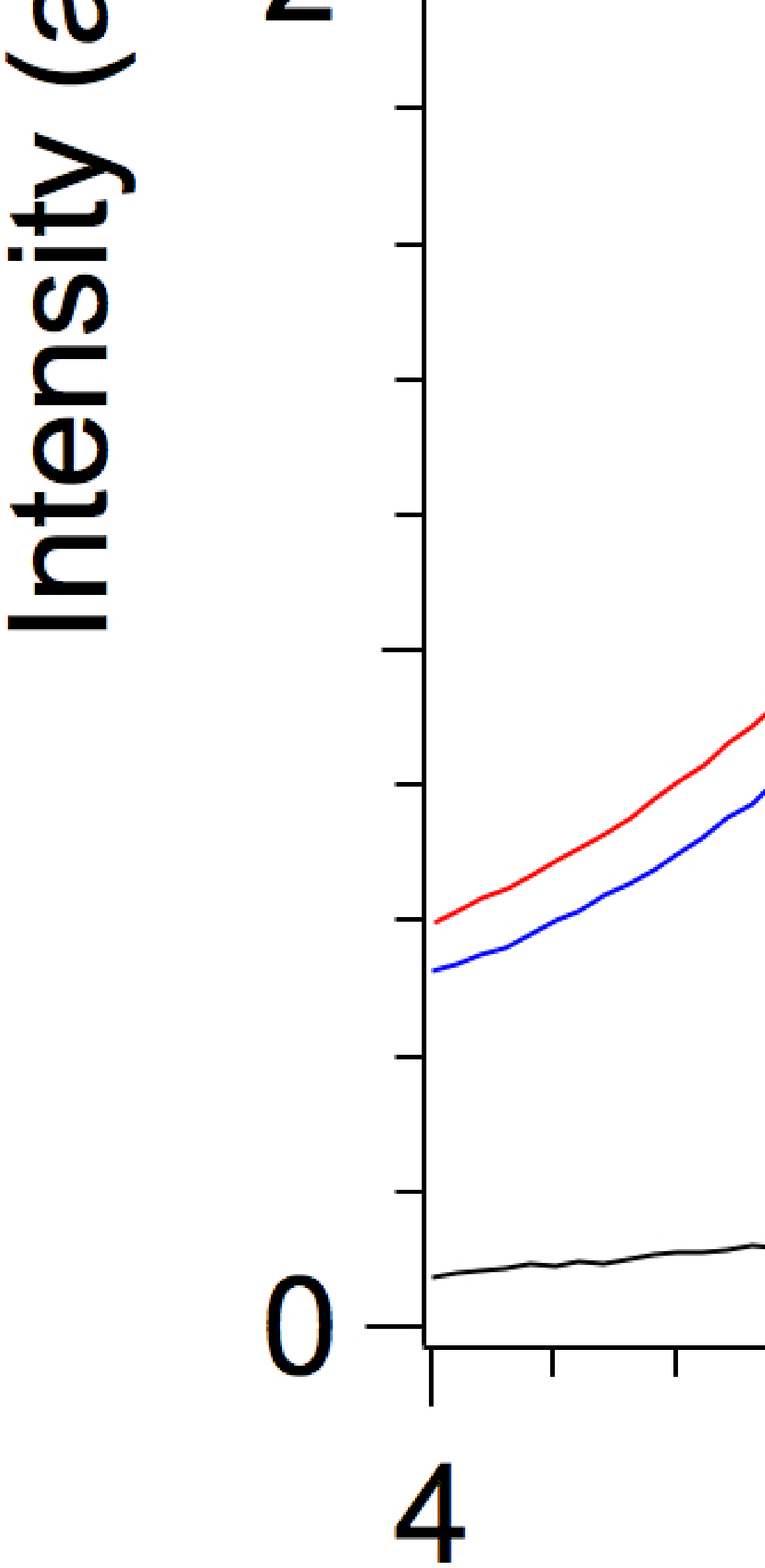}
\caption{\textbf{4\textit{f} flat bands in CeBi.} \textbf{a}, ARPES spectra of CeBi around the $\bar{\Gamma}$ point of the Brillouin zone at photon energy $h\nu$ = 122 eV, showing the existence of a 4\textit{f} flat band close to the Fermi level ($\sim$ 0.6 eV) and another 4\textit{f} flat band at deeper binding energies ($\sim$ 2.8 eV). \textbf{b}, same as \textbf{a} obtained at $h\nu$ = 128 eV. \textbf{c}, The difference of the two spectra in \textbf{a} and \textbf{b}. \textbf{d}, Momentum-integrated energy distribution curves (EDCs) at the photon energies of 122 eV and 128 eV and their difference.}
\end{figure}

\section{S\lowercase{lab bandstructure calculations}}\la{app:slab}

\begin{figure}[H]
\centering
\includegraphics[width=7.5cm]{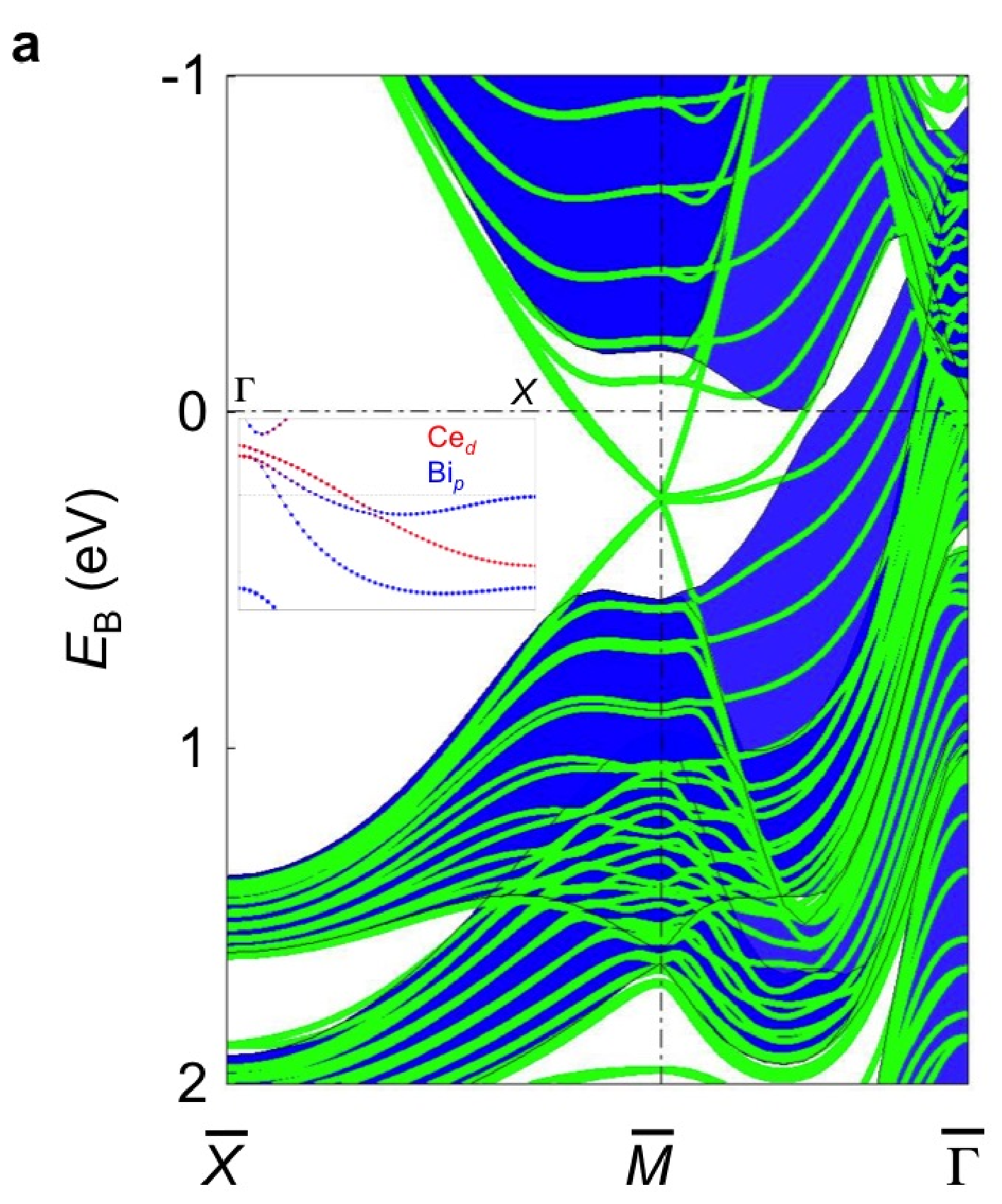}
\caption{\textbf{Calculation of CeBi slab.} Green lines: first-principles calculation of a CeBi 001 slab with a fine-tuned surface potential. Blue shaded areas correspond to bulk bands (from a first-principles bulk calculation) projected onto the surface Brillouin zone. The inset shows the inversion of Ce-$d$ and Bi-$p$ bands at the $X$ point of the bulk BZ. }
\end{figure}

\begin{figure}[H]
\centering
\includegraphics[width=16cm]{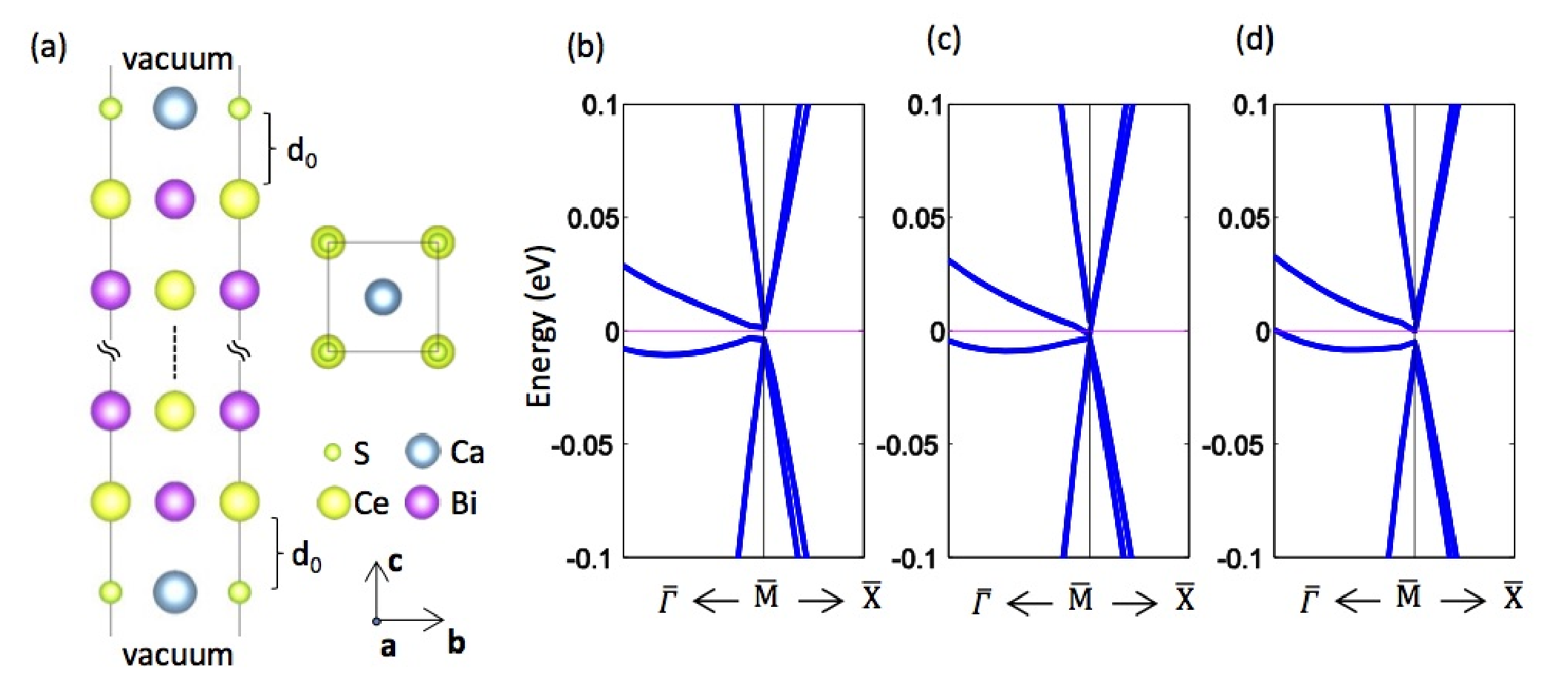}
\caption{\textbf{Details of the CeBi slab calculation.} The surface states of a 19-layer fccc  (001) nonmagnetic CeBi slab,  in which the surface Ce and Bi  atoms are replaced by Ca and S,  respectively. \textbf{a}, Top and side  views of  the slab geometry. \textbf{b-d}, The slab calculation  shows two strongly anisotropic and  near-degenerate Dirac cones at $\bar{M}$. The energy gap near zero energy depends on the spacing ($d_0$ in \textbf{a}) of the two outermost surface layers; for \textbf{b-d} respectively, $d_0=3.5,3.27$ and $3.03 \A$.}
\end{figure}

\begin{figure}[H]
\centering
\includegraphics[width=12cm]{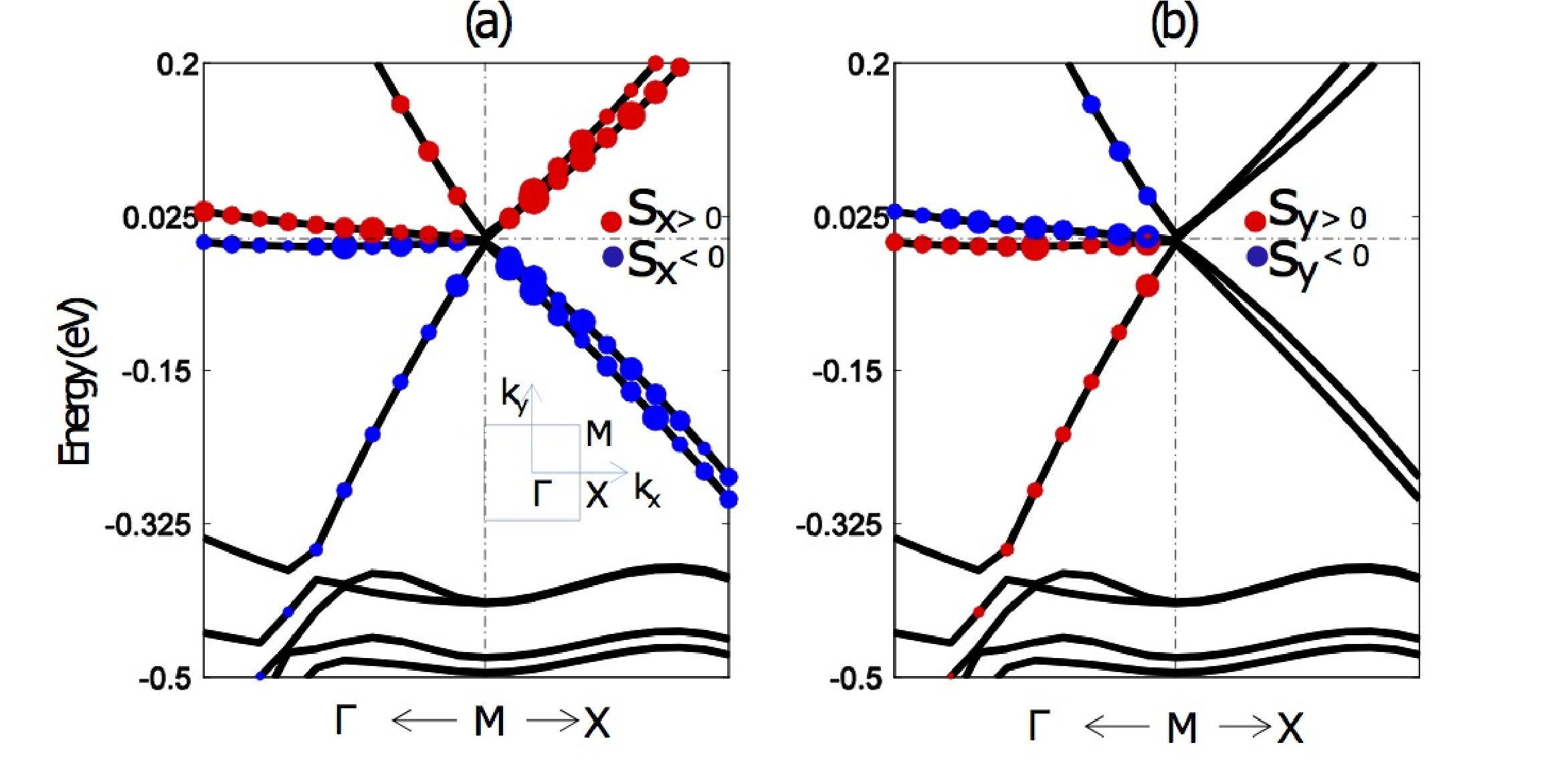}
\caption{\textbf{Rashba-type spin texture of CeBi surface states.} \textbf{a}, and \textbf{b}, show the expectation values of spin along $\vec{x}$ 
and $\vec{y}$ directions respectively.  For \textbf{a} and \textbf{b} the size of the dot indicates the magnitude of spin polarization, and the red (blue) color indicates that the spin is aligned in the positive (negative) direction. The Rashba-type texture is most evident along the high-symmetry line $M-X$, where spin (aligned in $\vec{y}$) and momentum (relative to $M$) are orthogonal.  }
\end{figure}

\section{E\lowercase{ffective Hamiltonian of the surface} D\lowercase{irac fermions}}\la{app:eff}

Our goal is to derive the effective Hamiltonian of the surface Dirac fermions:
\\
\bal \label{effhamsurf}
\calh(k_x,k_y) = \begin{pmatrix} v_1 k_x \sy - v_2 k_y \sx & 0 \\ 0 & v_2 k_x \sy - v_1 k_y \sx \end{pmatrix},  
\end{align}
\\
where $v_1$ and $v_2$ can be fitted to the experimental data of Ce\textit{X} (\textit{X} = Bi, Sb), and $\sigma_j$ are the Pauli matrices in a pseudospin representation. Our momentum coordinates $(k_x,k_y)$ are chosen relative to the high-symmetry point $\bar{M}$ in the 
(001) surface BZ, and their directions are parallel to the axes in Fig. 1\textbf{b} of the main text. The basis of $\calh$ are surface states with the orbital characters
\\
\bal \label{orbital}
&1: id_{yz}\uparrow + \frac{\eta}{\sqrt{2}}(-ip_y \uparrow + p_z\downarrow), \lin
&2: id_{yz}\downarrow + \frac{\eta}{\sqrt{2}}(-ip_y \downarrow + p_z\uparrow), \lin
&3: d_{xz}\uparrow + \frac{\eta}{\sqrt{2}}(p_x\uparrow+p_z\downarrow), \lin
&4: d_{xz}\downarrow +\frac{\eta}{\sqrt{2}}(p_x\downarrow-p_z\uparrow).
\end{align}
\\
Here, $d_{yz}$ and $d_{xz}$ correspond to Ce-$d$ orbitals, $p_x$, $p_y$, and $p_z$ to \textit{X}-$p$ orbitals (\textit{X} = Bi, Sb), $\uparrow$ refers to the spin component $S_z=1/2$, and $\eta \in \pm 1$ is the sign of $v_2$, which cannot be determined from the measured energy dispersion. The group of the wavevector $\bar{M}$ is the point group $C_{4v}$ \cite{C4v} combined with time-reversal symmetry ($T$), and its generators are represented by
\\
\bal \label{C4vgen}
&T = \begin{pmatrix} i\sy & 0 \\ 0 & -i\sy \end{pmatrix}K, \;\; C_{4z} = \begin{pmatrix} 0&0&-\lambda&0 \\ 0&0&0&\lambda^* \\ -\lambda&0&0&0 \\ 0&\lambda^*&0&0 \end{pmatrix} \ins{and} M_x = \begin{pmatrix} -i\sx & 0 \\ 0& i\sx \end{pmatrix}.
\end{align}
\\
Here, $C_{4z}$ is the four-fold rotation about $\hat{z}$, $M_x$ reflects $x \rightarrow -x$, $K$ implements complex conjugation and $\lambda=e^{i\pi/4}$.

We emphasize that $\calh$ is a minimal model of two unhybridized Dirac fermions, where each fermion originates from a bulk inversion at a different $X$ point. Indeed, the hybridization 
\\
\bal
\delta \calh = \Delta  \begin{pmatrix} 0 & \sz \\ \sz & 0 \end{pmatrix}
\end{align}
\\
preserves all the symmetries, and can in principle gap out the four-fold degeneracy where the two Dirac nodes overlap. However, choosing $\Delta=0$ and disregarding other symmetry-allowed hybridizations, our minimal model (Fig. 2\textbf{d} in the main text) closely overlaps with our measurements (Fig. 2\textbf{a}-\textbf{c}).

In the remainder of this section, we describe how exactly each surface Dirac fermion arises from a bulk inversion at an $X$ point. Specifically, the Dirac fermion in the upper block of $\calh$ derives from an inversion at $X_1=2\pi \hat{x}/a$, and the Dirac fermion in the lower block derives from an inversion at $X_2=2\pi \hat{y}/a$.  We begin with a low-energy description of $X_1$ involving only four bands: a Ce-$d$ doublet and a \textit{X}-$p$ doublet  (\textit{X} = Bi, Sb). Our effective Hamiltonian is
\\
\bal \label{H1}
H_{1}(k_x+2\pi/a,k_y,k_z) = \begin{pmatrix} \varepsilon_d &  v_1\sy k_x-v_2\sx k_y  +iv_2 k_z \\
                                    v_1\sy k_x -v_2\sx k_y  -iv_2 k_z &  \varepsilon_p \end{pmatrix},
\end{align} 
\\
where our momentum coordinates are relative to $X_1$, and our basis vectors are Bloch waves with orbital characters
\bal \label{basis1}
1: id_{yz}\uparrow, \;\;2: id_{yz}\downarrow, \;\; 3: -ip_y \uparrow + p_z\downarrow,\;\; 4:-ip_y\downarrow +p_z\uparrow;
\end{align}
these orbital characters are obtained from our first-principles calculation. We have introduced the bulk parameters $\varepsilon_d$, $\varepsilon_p$, $v_1$ and $v_2$; the last two parameters will be shown to coincide with that of the surface Hamiltonian ($\calh$). The form of $H_1$ can be deduced from knowledge of the group of the wavevector $X_1$. This is the point group $D_{4h}$ \cite{mildred} combined with time-reversal symmetry ($T$), and its generators are represented as
\\
\bal
T = \begin{pmatrix} i\sy & 0 \\ 0 & i\sy \end{pmatrix}&K, \;\; \cali = \matrixtwo{I}{0}{0}{-I},\;\; C_{4x} = \frac1{\sqrt{2}}\matrixtwo{-I+i\sigma_1}{0}{0}{-I-i\sigma_1}\lin
&M_y= \begin{pmatrix} i\sy & 0 \\ 0 & i\sy \end{pmatrix} \ins{and} M_x = \begin{pmatrix} -i\sx & 0 \\ 0& -i\sx \end{pmatrix}.
\end{align}
\\
Here, $\cali$ is the inversion operator, $C_{4x}$ implements four-fold rotation about $\hat{x}$, and $M_y$ reflects $y \rightarrow -y$. Following the notation in Ref. \onlinecite{kosterspacegroups}, the Ce-$d$ doublet (upper block of $H_1$) transforms in the $M_7^+$ representation of Tab. LIII, while the \emph{X}-$p$ doublet (lower block) transforms in the $M_7^-$ representation.

Up to a constant offset, $H_1$ describes a 3D Dirac fermion with mass $m=(\varepsilon_d-\varepsilon_p)/2$. This mass is inverted for CeSb (resp. CeBi), i.e., $\varepsilon_d<\varepsilon_p$, as illustrated in Fig. 4(a) (resp. Fig. S8). To derive the surface state that corresponds to this inversion, we model an interface between CeSb and vacuum by a kink in this mass parameter \cite{field1, field2, field3}, i.e., we let $m(z)$ depend on the spatial coordinate $z$, such that $m(z)=(\varepsilon_d-\varepsilon_p)/2<0$ for $z$ within CeSb, and $m(z)$ is large and positive for $z$ in the vacuum. Let us show that gapless states localize on this mass kink. In the envelop-function approximation, we replace $k_z \rightarrow -i\hbar \partial_z$ in $H_1$, and solve for the eigenfunctions of $H_1(z,\partial_z)$:
\\
\bal
\psi_{k_x,k_y,\pm }(z) = e^{ik_xx+ik_yy}\rho(z)\begin{pmatrix} 1 \\ \pm(\hat{n}_x+i\hat{n}_y)\\ \eta \\ \pm \eta (\hat{n}_x+i\hat{n}_y) \end{pmatrix},
\end{align}
\\
where $\eta$ is the sign of $v_2$, and 
\\
\bal
\hat{n}_x = -\frac{v_2k_y}{\sqrt{v_2^2k_y^2+ v_1^2k_x^2}}, \;\; \hat{n}_y = \frac{v_1k_x}{\sqrt{v_2^2k_y^2+ v_1^2k_x^2}}.
\end{align}
\\
$\rho(z)$ is generically a function that is localized at the interface; if $m(z)$ is a step function, then $\rho(z)$ exponentially decays on both sides of the interface. The subscripts $\pm$ on $\psi$ label the two bands of a massless Dirac fermion:
\\
\bal
H_1(z,\partial_z)\;\psi_{k_x,k_y,\pm }(z)=\pm \text{sgn}[v_2] \sqrt{ v_1^2k_x^2+v_2^2k_y^2}\; \psi_{k_x,k_y,\pm }(z).
\end{align}
\\
In a more convenient basis given by Eq.\ (\ref{orbital}), $\psi_{\pm}$ diagonalizes the upper block of the Hamiltonian $\calh$; cf. Eq.\ (\ref{effhamsurf}). The lower block is derived similarly from the 3D massive Dirac fermion at $X_2$, which is described by
\\
\bal \label{H2}
H_{2}(k_x,k_y+2\pi/a,k_z) = \begin{pmatrix}  \varepsilon_d &  -v_2\sy k_x+ v_1\sx k_y  -iv_2 k_z \\
                                    -v_2\sy k_x+ v_1\sx k_y  +iv_2 k_z &  \varepsilon_p \end{pmatrix}
\end{align} 
\\
in the basis
\bal \label{basis2}
1: d_{xz}\uparrow, \;\;2: d_{xz}\downarrow, \;\;3:p_x \uparrow + p_z\downarrow,\;\;4: p_x\downarrow -p_z\uparrow.
\end{align}
To lowest order in momentum, $H_1$ and $H_2$ are described by the same four parameters: $\varepsilon_d$, $\varepsilon_p$, $v_1$ and $v_2$. This coincidence arises because $X_1=2\pi \hat{x}/a$ and $X_2=2\pi \hat{y}/a$ are related by a four-fold rotation about $\hat{z}$, which is a symmetry of the rocksalt. In a straightforward generalization of $X_1$, we similarly derive a \emph{second}  massless, surface Dirac fermion due to the bulk inversion at $X_2$; this second fermion is described by the lower block of $\calh$ in Eq.\ (\ref{effhamsurf}).

Finally, we describe the 3D massive Dirac fermion at $X_3=2\pi \hat{z}/a$, which is related to $X_1$ and $X_2$ by cubic symmetry. The low-energy Hamiltonian is
\\
\bal \label{H3}
H_{3}(k_x,k_y,k_z+2\pi/a) = \matrixtwo{\varepsilon_d}{-iv_2k_y -v_1 k_z\sx +v_2k_x\sz}{iv_2k_y -v_1 k_z\sx +v_2k_x\sz}{\varepsilon_p},
\end{align}
\\
in the basis
\bal \label{basis3}
1:d_{xy}\uparrow, \;\;2: d_{xy}\downarrow, \;\;3:(p_{x}+ip_{y})\uparrow,\;\;4:(p_{x}-ip_{y})\downarrow.
\end{align}
From the preceding discussion, one might expect a third massless Dirac fermion centered at $\bar{\Gamma}$ in the 001 BZ; note $\bar{\Gamma}$ is the projection of the bulk $X_3$ point, as illustrated in Fig. 1\textbf{b}. As explained in the main text, we expect this third surface cone to be masked by bulk bands.

\section{P\lowercase{ossible topologies under crystalline symmetries}}\la{app:poss}

Given that parity inversions at three $X$ points determine Ce$X$ to be a strong topological insulator in the time-reversal-symmetric classification, there remains a finer distinction of quantum groundstates which is afforded by crystalline symmetries. This finer distinction  may be formalized by an integer invariant called the mirror Chern number \cite{teo} ($\calc_+$). Our goal is to evaluate $\calc_+$ in the plane $k_y=0$, which is invariant under the reflection $M_y: y \rightarrow -y$. This plane projects onto a mirror line which intersects $\bar{\Gamma}$ and $\bar{M}$ in the 001 BZ, as illustrated in Fig. 1\textbf{b} of the main text.  Recalling the definition of $v_1$ and $v_2$ as parameters of the low-energy Hamiltonians at three bulk $X$ points (see Eq.\ (\ref{H1}), (\ref{H2}) and (\ref{H3})), we define $\chi$ as the relative sign of $v_1$ and $v_2$. $\chi$ is a $\Z_2$ parameter that distinguishes between two types of topological crystalline insulators: if $\chi=+1$ (resp. $-1$), we find that $\calc_+=+1$ (resp. $-3$).

We define $\calc_+$ (resp. $\calc_-$) as the integrated Berry flux of occupied, mirror-even (resp. -odd) bands in the $k_y=0$ plane \cite{teo}. By mirror-even (resp. -odd), we mean that the band is an eigenstate of reflection ($M_y$) with eigenvalue $+i$ (resp. $-i$). Time-reversal symmetry enforces that $\calc_+ =-\calc_-$, i.e., there is only one independent integer invariant. By `occupied' bands, we also include the bulk bands that rise above the Fermi level around the bulk $\Gamma$ point; only by this inclusion can $\calc_+$ be quantized to integers.

The integrated Berry flux has three contributions which can be independently determined at each $X$ point. We focus first on $X_1$, where we rewrite Eq.\ (\ref{H1}) as
\bal
H_{1}(k_x+2\pi/a,0,k_z) = m\tau_3 + v_1 k_x \tx \sy -v_2 k_z \ty
\end{align}
by fixing $k_y=0$ and substracting the constant-energy offset. Here, we have also introduced a second set of Pauli matrices where $\tau_3=+1$ corresponds to the upper two-dimensional block in Eq.\ (\ref{H1}); $\tx \sy$ is shorthand for $\tx \otimes \sy$, and $\tz$ for $\tz \otimes$ (identity in the $\sigma$ space). From the basis of $H_1$ in Eq.\ (\ref{basis1}), we deduce the representation of reflection as $M_y = i\sy$. In each mirror subspace with eigenvalue $i\kappa$ ($\kappa =\pm 1)$, we thus have a two-dimensional Hamiltonian which we label by a superscript $\kappa$:
\bal
H_1^{\kappa}(k_x+2\pi/a,0,k_z) = m\tau_3 + v_1 \kappa k_x \tx -v_2 k_z \ty.
\end{align}
By rewriting $H_1^{\kappa} = \sum_{i,j=1}^3 K_i [A_1^{\kappa}]_{ij} \tau_j$ with $K_1=k_x, K_2=k_z$ and $K_z=m$, we define a useful quantity $\calc_1^{\kappa}$ as the sign of the determinant of $A_1^{\kappa}$; here, $\calc_1^{\kappa} = -\chi \kappa$. $-\calc_1^{+}$ is the change in Berry flux of the occupied, mirror-even bands due to a band touching at $X_1$, where the mass ($m$) inverts sign \cite{BABtextbook}. By summing $\calc_1^+$ with contributions (denoted $\calc_2^+$ and $\calc_3^+$) from the other band touchings at $X_2$ and $X_3$, we determine the net change of the Berry flux for the entire mirror plane, i.e.,  the mirror Chern number $\calc_+ = -\calc_1^+-\calc_2^+-\calc_3^+$. In the remainder of this Section, we proceed to evaluate $\calc_2^+$ and $\calc_3^+$.

To evaluate $\calc_2^+$, we first address a subtlety that $X_2=2\pi \hat{y}/a$ does not lie on the mirror plane $k_y=0$, but $X_2 + G$ does, with $G= 2\pi (\hat{x}-\hat{y}+\hat{z})/a$ a reciprocal lattice vector. The Hamiltonians $H(k)$ and $H(k+G)$ are related by a gauge transformation which reflects the aperiodicity of our Bloch-wave basis in a multi-atomic Bravais lattice \cite{aaWilson}.  After accounting for this gauge transformation, the low-energy Hamiltonian about $X_2+G$ is 
\\
\bal \label{H2trans}
{H}_{2}(k_x+\tfrac{2\pi}{a},k_y,k_z+\tfrac{2\pi}{a}) = \begin{pmatrix} \varepsilon_d &  v_2\sy k_x- v_1\sx k_y  +iv_2 k_z \\
                                    v_2\sy k_x-v_1\sx k_y  -iv_2 k_z &  \varepsilon_p \end{pmatrix}.
\end{align}
\\
By substracting the constant-energy offset, this can be rewritten as
\bal
H_2(k_x+\tfrac{2\pi}{a},0,k_z+\tfrac{2\pi}{a}) =	m\tau_3 + v_2 k_x \tx \sy -v_2k_z \ty,
\end{align}
where, once again, $\tau_3=+1$ corresponds to the upper two-dimensional block in Eq.\ (\ref{H2trans}). The representation of reflection in the basis of $H_2$ is $M_y=-i\sy$ (from Eq.\ (\ref{basis2})), which crucially differs in sign from the representation at $X_1$. In a mirror subspace labelled by eigenvalue $i\kappa$, we then have
\\
\bal
H_2^{\kappa}(k_x+\tfrac{2\pi}{a},0,k_z+\tfrac{2\pi}{a}) =	m\tau_3-\kappa v_2 k_x \tx -v_2k_z \ty = \begin{pmatrix} k_x\; k_z\; m \end{pmatrix}\begin{pmatrix} -\kappa v_2 & 0 & 0 \\ 0& -v_2 & 0 \\ 0&0& 1 \end{pmatrix} \begin{pmatrix} \tau_1 \\ \tau_2 \\ \tau_3 \end{pmatrix}.
\end{align}
\\
$\calc_2^{\kappa}$ is the sign of the determinant of the above matrix, and equals $\kappa$. 

Finally at $X_3$, we have that
\bal
H_3(k_x,0,k_z+\tfrac{2\pi}{a}) = m\tau_3 -v_1k_z \tx \sx +v_2k_x\tx\sz,
\end{align}
where the reflection is represented in this basis (cf. Eq.\ (\ref{basis3})) by $M_y = i\tz\sy$. A new basis may be found where in each mirror subspace,
\\
\bal
H^{\kappa}_3(k_x,0,k_z+\tfrac{2\pi}{a}) = m\gamma_3 +v_2k_x\gamma_1 -\kappa v_1 k_z \gamma_2 = \begin{pmatrix} k_x\; k_z\; m \end{pmatrix}\begin{pmatrix} v_2 & 0 & 0 \\ 0& -\kappa v_1 & 0 \\ 0&0& 1 \end{pmatrix} \begin{pmatrix} \gamma_1 \\ \gamma_2 \\ \gamma_3 \end{pmatrix},
\end{align}
\\
with $\gamma$ another set of Pauli matrices. We thus derive $\calc_3^{\kappa} = -\chi \kappa$.

Summing all three contributions,
\bal
\calc_{\kappa} = -\calc_1^{\kappa}-\calc_2^{\kappa}-\calc_3^{\kappa} = \kappa (2 \chi -1).
\end{align}
We have thus proven our claim that $\calc_+ = +1$ (resp. $-3$) if $\chi=+1$ (resp. $-1$). We may restate this result in terms of surface bands. Along the $k_y=0$ mirror line in the 001 BZ (cf. Fig. 1\textbf{b} in the main text), our analysis predicts one chiral surface band for each bulk inversion, with an anti-chiral partner having opposite mirror eigenvalue. The bulk inversion at $X_1$ leads to the partnered surface dispersions  
\bal
E^{\kappa}_1(k_x) = \chi \kappa |v_1| k_x,
\end{align}
where once again $\kappa=+1$ for mirror-even bands, and $k_x$ is centered at $\bar{M}$. Similarly for $X_2$,
\bal
E^{\kappa}_2(k_x) =  - \kappa |v_2| k_x.
\end{align}
Here, $E^{\kappa_1}_1(0)=E^{\kappa_2}_2(0)$ for the four combinations of $(\kappa_1,\kappa_2)$ restates the four-fold degeneracy at $\bar{M}$, if there is no hybridization between the two flavors of fermions. If $\chi=+1$, there are two chiral modes centered at $\bar{M}$ with opposite mirror eigenvalues -- such surface states can be fully gapped while preserving the symmetries. On the other hand, if $\chi=-1$, the two chiral modes have the same mirror eigenvalues -- while the four-fold degeneracy at $\bar{M}$ can be lifted in a symmetric way, the surface states around $\bar{M}$ cannot be fully gapped.

To conclude, we consider the surface bands expected from a bulk inversion at $X_3$: 
\bal
E^{\kappa}_3(k_x) = \chi \kappa |v_2| k_x
\end{align}
for $k_x$ centered around $\bar{\Gamma}$. Once again, these last surface bands are effectively masked by bulk bands.

\newpage

\noindent\textbf{Supplementary References}